\documentclass[letterpaper,twocolumn,10pt]{article}
\usepackage{usenix-2020-09}

\usepackage{listings}
\usepackage{lstautogobble}
\usepackage[acronym]{glossaries}
\usepackage[inline]{enumitem}
\usepackage{ifxetex,ifluatex}
\usepackage[utf8]{inputenc}
\usepackage{subcaption}
\usepackage{amsthm}
\usepackage{graphicx}
\usepackage{amsmath,amssymb}

\theoremstyle{definition}
\newtheorem*{authprop}{Authentication Property}
\newtheorem*{privprop}{Privacy Property}

\glsdisablehyper{}
\newacronym{uri}{URI}{uniform resource identifier}
\newacronym{dns}{DNS}{domain name system}
\newacronym[longplural={certificate authorities}]{ca}{CA}{certificate authority}
\newacronym{pki}{PKI}{public key infrastructure}
\newacronym{cdn}{CDN}{content delivery network}
\newacronym{json}{JSON}{JavaScript Object Notation}
\newacronym{ct}{CT}{Certificate Transparency}
\newacronym{kdc}{KDC}{key distribution center}
\newacronym{acme}{ACME}{Automatic Certificate Management Environment}
\newacronym{rfc}{RFC}{Request for Comments}
\newacronym{ietf}{IETF}{Internet Engineering Task Force}
\newacronym{mrs}{MRS}{multi-set rewriting system}
\newacronym{AnB}{AnB}{Alice-and-Bob}
\newacronym{otp}{OTP}{one-time password}
\newacronym{idp}{IdP}{identity provider}
\newacronym{jws}{JWS}{JSON Web Signature}
\newacronym{pkce}{PKCE}{Proof Key for Code Exchange}
\newacronym{rp}{RP}{relying party}
\newacronym{csrf}{CSRF}{cross-side request forgery}
\newacronym{wim}{WIM}{Web Infrastructure Model}
\newacronym{mitm}{MITM}{Meddler-in-the-Middle}
\newacronym{fapi}{FAPI}{OpenID Financial-grade API}
\newacronym{saas}{SaaS}{Software-as-a-Service}
\newacronym{sso}{SSO}{Single Sign-On}
\newacronym{poidc}{POIDC}{Privacy-Preserving OpenID Connect}

\def\anonymize{0}

\title{\Large \bf SOAP: A Social Authentication Protocol}
\author{
  \if\anonymize0
  {\rm Felix Linker}\\
  Department of Computer Science, ETH Zurich
  \and
  {\rm David Basin}\\
  Department of Computer Science, ETH Zurich
  \fi
}

\date{}

\usepackage{biblatex}
\begin{document}

\maketitle

\begin{abstract}
  Social authentication has been suggested as a usable authentication ceremony to replace manual key authentication in messaging applications.
  Using social authentication, chat partners authenticate their peers using digital identities managed by \acrlongpl{idp}.
  In this paper, we formally define social authentication, present a protocol called SOAP that largely automates social authentication, formally prove SOAP's security, and demonstrate SOAP's practicality in two prototypes.
  One prototype is web-based, and the other is implemented in the open-source Signal messaging application.

  Using SOAP, users can significantly raise the bar for compromising their messaging accounts.
  In contrast to the default security provided by messaging applications such as Signal and WhatsApp, attackers must compromise both the messaging account and all \acrlong{idp}-managed identities to attack a victim.
  In addition to its security and automation, SOAP is straightforward to adopt as it is built on top of the well-established OpenID Connect protocol.
\end{abstract}

\section{Introduction}
\textit{Social authentication} promises simple, usable, and remote key authentication for messaging applications \cite{SocialAuthentication} and was first implemented in the Keybase application \cite{Keybase}.
Using Keybase, Alice can link her Keybase account to, for example, her Twitter account by tweeting a message signed with her Keybase account's key.
This allows other users to \textit{socially authenticate} Alice on Keybase via her Twitter account.
More generally, when performing social authentication, users verify that their actual chat partner controls accounts at different \glspl{idp} which they know are controlled by their intended chat partner.

Authenticating chat partners is critical for user security: if not done properly, users risk that a \gls{mitm} intercepts their messages.
Existing authentication ceremonies do not sufficiently address this risk.
Various studies have found that users are unwilling or unable to perform these authentication ceremonies \cite{E2EAuthUserStudy3,E2EAuthUserStudy1,RemoteFingerprintComp,E2EAuthUserStudy2}.
In particular, users are both challenged and constrained by the in-person comparison of safety numbers as implemented in the messaging applications Signal and WhatsApp.
Not only must they understand how to perform this ceremony correctly, they must also be in close physical proximity with one another.

\begin{figure}[t]
  \centering
  \includegraphics[width=\columnwidth]{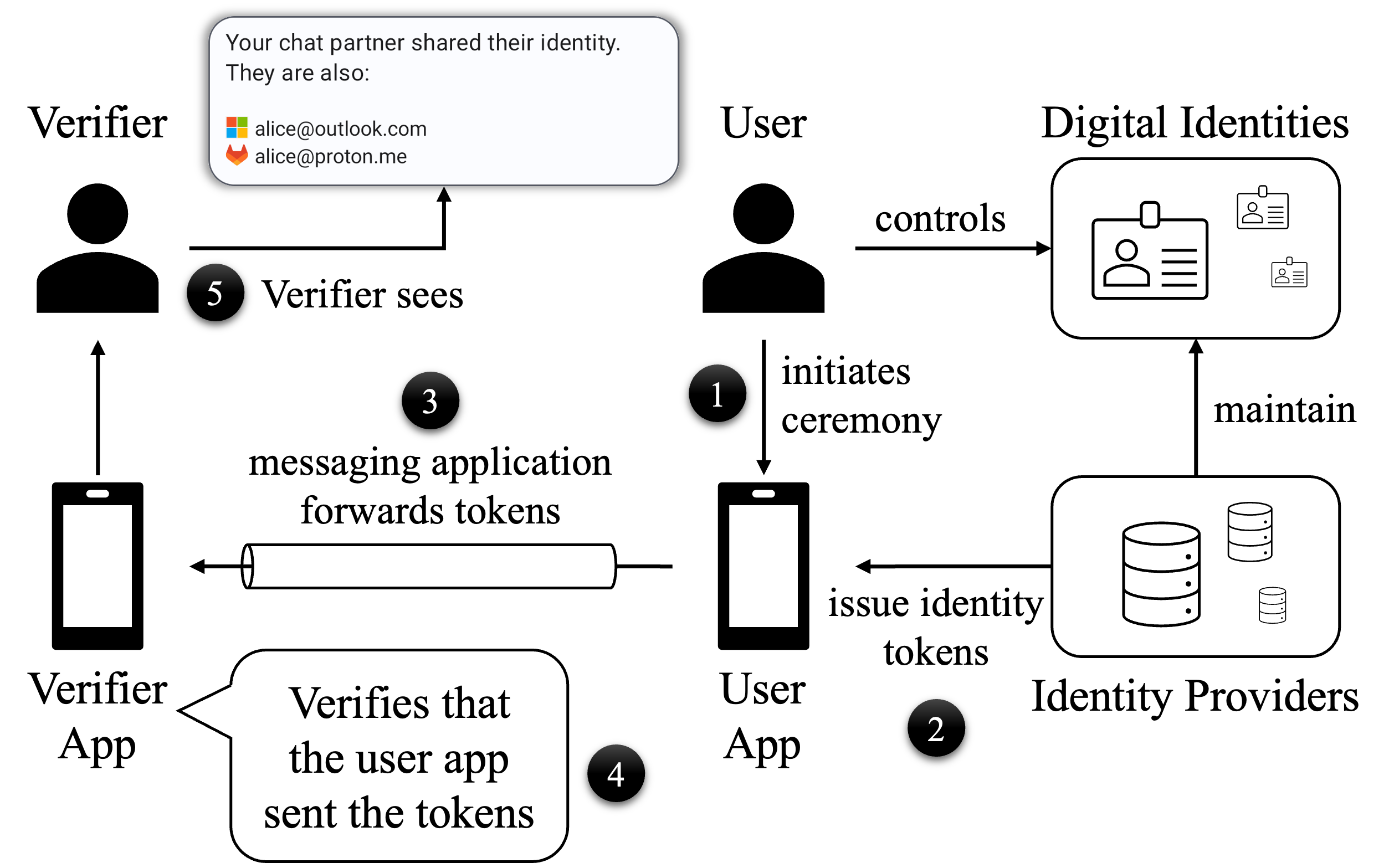}
  \caption{SOAP implements a social authentication ceremony.
  A user initiates the ceremony in their messaging application, which requests an identity token for each of the user's identities and forwards the tokens.
  The verifier's application verifies the token's sender.
  The verifier uses the identities to authenticate the user.}
  \label{fig:overview}
\end{figure}

In contrast, automated social authentication was established as a usable authentication ceremony \cite{SocialAuthentication} that works remotely.
Keybase was first to study social authentication beyond the idea, but Keybase requires manually posting key material, which requires non-trivial user effort.
Moreover, the posting is public, which discloses account associations to everyone.

After Zoom acquired Keybase, Zoom published an end-to-end encryption whitepaper \cite{ZoomE2EWhitepaper} which continued this line of work.
In particular, it automated the authentication process using a modified version of the OpenID Connect protocol.
Zoom's proposal, though, was designed in a setting where every account can be authenticated only by a single \gls{idp} and, moreover, Zoom's design requires \gls{idp} adoption.

Finally, although past works have presented designs, social authentication has never been studied as an authentication protocol.
No prior work defined what security guarantees social authentication should provide, let alone considered whether a given design correctly provides these guarantees.

In this paper, we address all these shortcomings and present SOAP, a formally verified social authentication protocol.
We make the following contributions:
\begin{itemize}[noitemsep]
  \item We precisely define the security objectives of social authentication and argue that it should provide a novel security property that we call \textit{sender correspondence}.
  This is a strong security property in that messaging sessions can only be compromised if all digital identities and the application's key servers are compromised.
  This raises the bar for the adversary and distributes trust amongst many providers.
  In contrast to Signal's and WhatsApp's default security, neither the cellular provider nor the key servers nor any of the \glspl{idp} involved can individually intercept messaging sessions.
  \item We formally relate sender correspondence to existing notions of authentication, and show how sender correspondence applies to designs beyond secure messaging.
  \item We present SOAP, a secure and practical protocol implementing social authentication.
  Our protocol can be seen as an extension of Zoom's design that works without \gls{idp} adoption and for multiple \glspl{idp}.
  SOAP automates the authentication ceremony and provides a straightforward and immediate means for adoption.
  Figure~\ref{fig:overview} provides an overview of our design.
  \item Using the Tamarin model checker \cite{Tamarin}, we formally prove that SOAP satisfies our novel security property and that SOAP respects user privacy.
  Users can decide to whom they disclose which identities, and SOAP leaks no information to \glspl{idp} beyond that one is using the messaging app in question, e.g., Signal.
  By employing a salt-and-hash scheme, we avoid revealing key material to \glspl{idp} and, thus, leaking one's contacts to providers.
  \item We show that SOAP is straightforward to adopt by implementing it in two fully functional prototypes: a web-based application and an extension of the Signal Android application.\footnote{%
    The web-based prototype is hosted at \url{https://soap-proto.net}, and a video demo of the Signal prototype can be viewed at \url{https://youtu.be/Ip_RAF8PRrM}.
  }
  The former requires some user interaction whereas the latter functions mostly automatically.
\end{itemize}

To the best of our knowledge, SOAP is the first formally verified authentication ceremony for messaging applications that works remotely and does not require users to work with cryptographic objects like keys or fingerprints.
By leveraging an existing and widely used standard, SOAP is easy to implement and can be used immediately with any \gls{idp} that already supports OpenID Connect.
Finally, SOAP allows for authentication in useful ways, that are impossible using conventional solutions, which we explicate in two further use cases.

\paragraph{Use Case 1: Social Authentication as a Second Factor}
One may question SOAP's value whenever users cannot authenticate their chat partner's identities because they have no relationship to these identities.
For example, someone who has never received an e-mail from a given Outlook address would be unable to verify that e-mail address as truly belonging to the person in question without further interaction.

In such cases, SOAP is still valuable as it can serve as a second factor, raising the bar for compromise.
If one of your contacts authenticated themselves as in control of two accounts, and you are prompted that this contact's public key changed, you can check whether the ``new'' contact still controls both these accounts.
This information can help to distinguish a public key maliciously associated to your contact's profile from a legitimate, fresh public key after key-rollover.

\paragraph{Use Case 2: Native Digital Authentication}
For some online interactions, users do not base the identification of their chat partners in the physical world, but rather in the digital world.
For example, in the physical world, I might like to authenticate my chat partner as ``my colleague Alice, who I eat lunch with every day.''
In contrast, in the digital world, I might like to authenticate my chat partner as ``the open-source maintainer Alice123 on GitHub, who I have never met in real life, but writes beautiful JavaScript.''
In the latter case, SOAP promises to seamlessly bootstrap a secure communication channel from such a pre-existing relationship.

\paragraph{Structure}
We proceed with additional problem motivation in Section~\ref{sec:bg}.
In Section~\ref{sec:prop}, we present SOAP's design idea, define our security goal, and provide our threat model.
Section~\ref{sec:design} explains SOAP's design, Section~\ref{sec:security} presents our security analysis, and Section~\ref{sec:proto} reports on our two prototypes.
Finally, Section~\ref{sec:rw} compares SOAP to related work and sender correspondence to existing notions of authentication and designs.

\paragraph{Artifacts}
Our formal model and proofs (Sec.~\ref{sec:security}), and prototypes (Sec.~\ref{sec:proto}) are available at: \url{https://soap-wg.github.io/sources}.
Both prototypes come with source files and compiled versions, e.g., .apk files for the Signal prototype.

\section{Problem Motivation}
\label{sec:bg}
\begin{figure}
  \centering
  \includegraphics[width=\columnwidth]{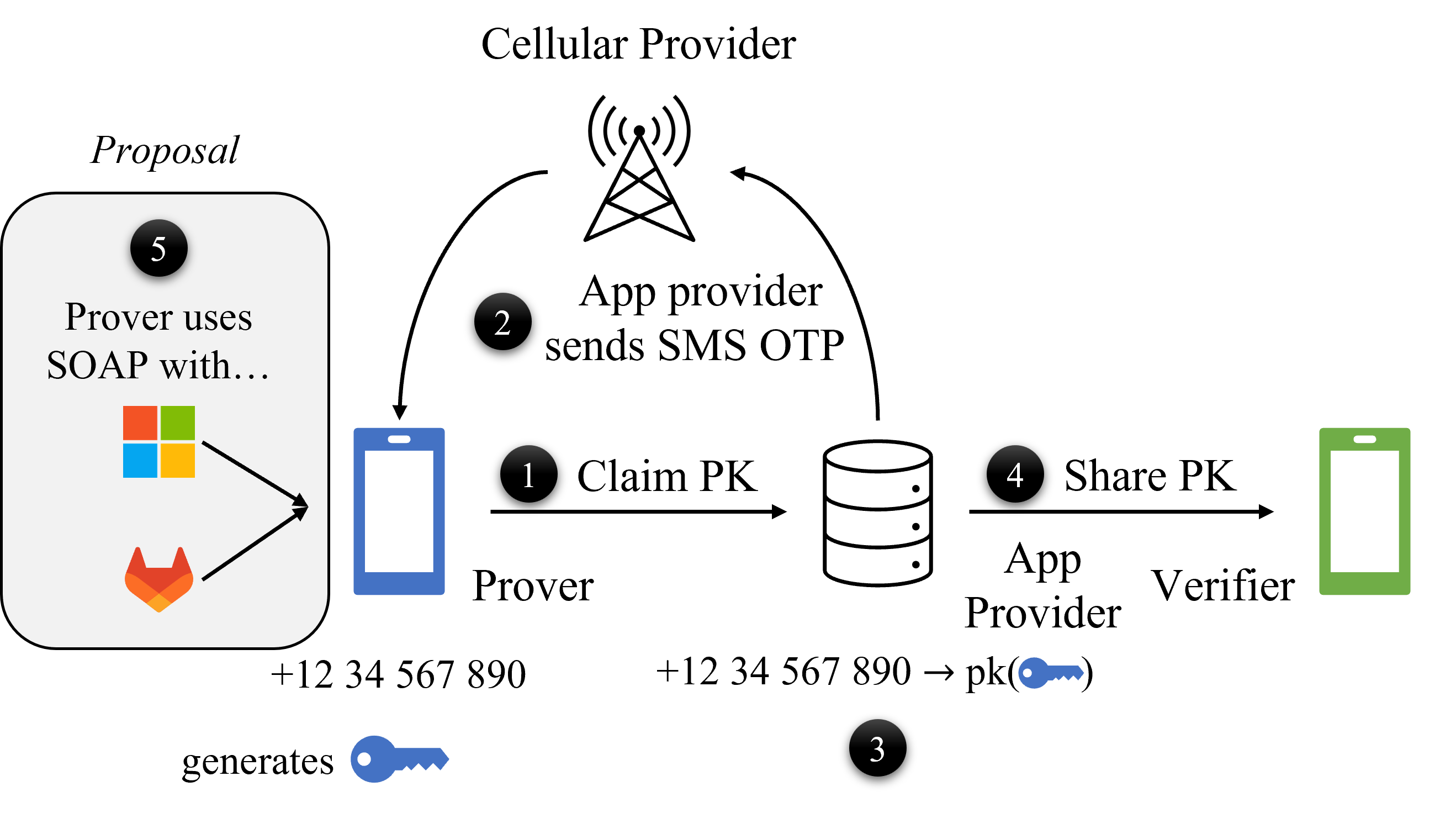}
  \caption{
    Steps of the prover to register at the messaging application and share their public key with other users.
    Our proposal is to additionally authenticate using multiple \glspl{idp}, here Microsoft and GitLab.
  }
  \label{fig:app-setup}
\end{figure}

Both Signal and WhatsApp allow their users to communicate without authenticating their chat partners.
By default, users rely on the authentication performed by the application provider during registration and that the application's key server correctly reports other users' public keys to them.
When registering, users first register a public key at the key server (Step~1, Fig.~\ref{fig:app-setup}).
For that, they generate a public/private key pair locally and claim that the respective public key is associated with their phone number.
The application then sends the user an SMS \gls{otp} to verify that the user controls the phone number (Step~2).
When the user enters the correct \gls{otp} into the app, the key server associates the user's public key with their phone number (Step~3).
Afterwards, the Signal server shares the user's public key with other users (Step~4).

This key distribution practice requires a high degree of trust in the security of SMSes and the application servers themselves.
An adversary can eavesdrop or impersonate a user by compromising that user's cellular provider \textit{or} the key server.
More specifically, SMS-based attacks like SIM swaps \cite{SMSAuthInsecure} allow for impersonation,  and by compromising a key server an adversary can eavesdrop on users as a \gls{mitm}.
Users must trust their messaging application and thus their messaging provider in general.
But it is one thing to trust a provider to not implement a backdoor in a messaging application, and another to trust that their key servers are never compromised by insiders, attackers, or force by government authorities.

The breach of Signal's SMS \gls{otp} provider Twilio \cite{SignalOTPCompromise} calls into question whether the considerable trust placed in Signal's registration procedure is warranted.
Through social engineering, attackers gained read access to SMS \glspl{otp}, and re-registered phone numbers, one of which belonged to a prominent journalist \cite{SignalOTPCompromiseVictim}.

To prevent such attacks, Signal and WhatsApp users can compare \textit{safety numbers} \cite{SignalSafetyNumbers,SignalWhatsAppE2E} with their chat partners.
Safety numbers are fingerprints of the two conversation participants' public keys.
If two chat partners' safety numbers match, they are using the same public keys, and hence there is no \gls{mitm}.
To compare safety numbers though, users must rely on a trusted out-of-band channel, for example, the in-person scanning of the QR codes that encode these numbers.

In Signal's Android app, the screen for safety number verification displays the safety number as a QR code and numerically.
Signal users have four options:
\begin{enumerate*}[label=(\roman*)]
  \item They can mark the conversation as verified.
  \item They can tap the QR code to scan their partner's QR code.
  \item They can share the safety number using a separate out-of-band channel.
  To do this, they click on the appropriate button, which launches the sharing menu of the phone's operating system.
  \item Finally, they can press on the safety number itself to either copy it to the clipboard or to compare it with the clipboard's contents.
\end{enumerate*}

Although safety numbers provide a secure authentication ceremony, their actual benefit is questionable.
In incidents like the Twilio breach, users are unlikely to reauthenticate their session in a timely way by scanning QR codes.
Signal notified the impersonated journalist's contacts when the attackers associated a malicious public key with the journalist's phone number.
However, to successfully thwart this attack, users must
\begin{enumerate*}[label=(\roman*)]
  \item notice and understand this warning, and
  \item compare safety numbers in-person, which requires physical proximity, or
  \item compare safety numbers using an ad-hoc out of band channel, which must be first agreed upon -- while possibly talking to an attacker.
\end{enumerate*}
But even if users were to engage in these authentication ceremonies, various studies suggest that an attack is still very likely to succeed.

To start with, \cite{E2EAuthUserStudy3,E2EAuthUserStudy1,E2EAuthUserStudy2} reported that, lacking explicit instructions, only around 15\%-25\% of user study participants manage to successfully authenticate their chat partners in person or using phone calls.
When receiving instructions, these numbers rise to 75\%-80\%.
Remarkably, only around 50\% of \cite{E2EAuthUserStudy3}'s participants indicated that they would perform the authentication ceremony again in the future, even after its importance was explained to them.
Another study investigated the usability of SMS-based authentication ceremonies supported by Signal's share button in the safety screen's top right corner \cite{RemoteFingerprintComp}.
The authors found that in 40\%-60\% of the cases, a difference in safety numbers went unnoticed.
Notably, the study considered a best-case scenario by recruiting educated, technology-savvy users who were explicitly instructed to compare safety numbers.

The aforementioned user studies show that many users cannot correctly compare safety numbers and, more critically, they do not want to.
To make matters even worse, none of these studies accounts for how few users initiate these ceremonies in the first place.
The share of authenticated messaging sessions is likely much lower than the above figures suggest.

Users can also protect their own accounts using a \textit{registration lock}, provided by both WhatsApp and Signal \cite{WhatsAppRegistrationLock,SignalRegistrationLock}.
When activating the registration lock, users configure a PIN or password that must be entered when attempting to associate a new public key with a phone number.
This, for the most part, rules out SMS-based attacks.
If users (or attackers) cannot provide the PIN, an inactivity of 7 days is required to disable the registration lock.
Registration locks, though, still require trust in the application's key server, and, more critically, users cannot verify whether their chat partners utilize them.
I can defend myself against impersonation attacks on my identity, but I cannot know whether my peers are impersonated or \gls{mitm}-ed.

\section{A Social Authentication Protocol}
\label{sec:prop}
To address all aforementioned issues of modern messaging applications, we propose SOAP and present its security goals, its design idea, and our threat model.

\subsection{Design Goals}

\begin{figure}
  \centering
  \includegraphics[width=.7\columnwidth]{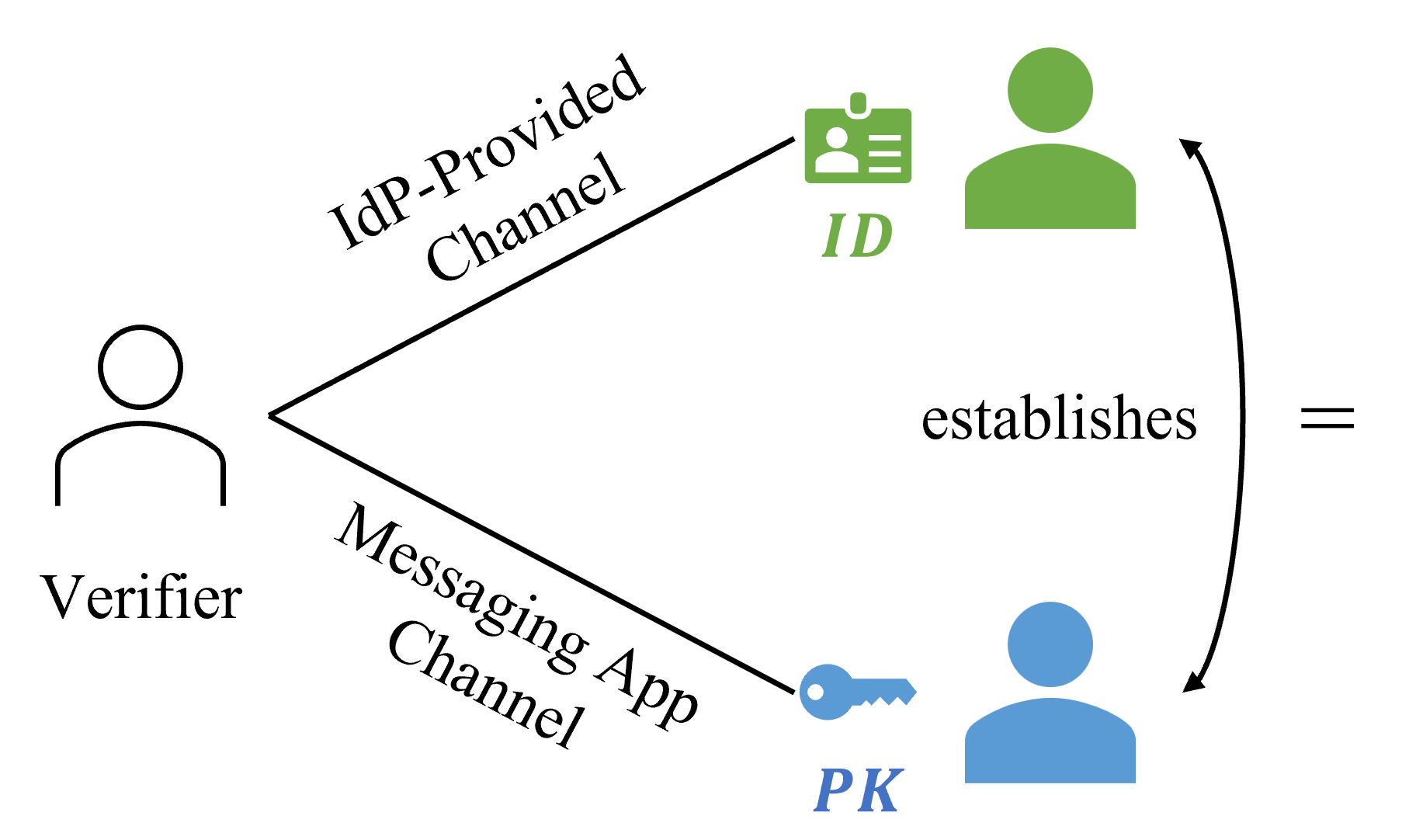}
  \caption{Social authentication establishes for a verifier that the digital identity $ID$ and messaging application public key $PK$ are controlled by the same person.}
  \label{fig:soc-auth}
\end{figure}

The aim of social authentication is for a verifier to authenticate a prover (the verifier's chat partner) using one or more digital identities.
Social authentication is appealing as:
\begin{enumerate*}[label=(\roman*)]
  \item many users have pre-existing relationships on social media, and
  \item by linking their social media presence to a different account, they can transfer the relationship from one medium to another.
\end{enumerate*}
Note that these two notions are independent: even if you have no relationship with a given social media account, you could still be convinced that you are talking to the account holder.
Formalizing this intuition, we define the security property of sender correspondence which, as we shall later see, is an authentication property.

\begin{authprop}[Sender correspondence]
  A protocol $P$ guarantees a verifier \textit{sender correspondence} between two pseudonyms $A$ and $B$ if, whenever $P$ successfully terminates, then all messages that appear to have been sent by $A$ and all messages that appear to have been sent by $B$ were sent by the same user.
\end{authprop}

Social authentication is simply an instantiation of sender correspondence, where the pseudonym $A$ is the messaging application public key $PK$ and $B$ is an \gls{idp}-controlled digital identity $ID$.
We define sender correspondence in more general terms than social authentication because sender correspondence finds application in other protocols, as we will discuss in Section~\ref{sec:rw-props}.
We illustrate social authentication's security guarantees in Figure~\ref{fig:soc-auth}.

In addition to social authentication, SOAP was also designed to provide privacy.
For example, \glspl{idp} cannot learn with whom their users communicate.
We define SOAP's privacy property in terms of the allowed leakage to an \gls{idp}.
In particular, \glspl{idp} neither learn who the prover authenticates to nor which other \glspl{idp} the prover uses.

\begin{privprop}
  \Glspl{idp} can only learn that SOAP users
  \begin{enumerate*}[label=(\roman*)]
    \item use SOAP,
    \item the messaging applications where they use it, and
    \item when they use it.
  \end{enumerate*}
\end{privprop}

\subsection{Design Idea}
\label{sec:idea-and-props}

SOAP, our Social Authentication Protocol, works as follows.
The prover requests an OpenID Connect identity token at an \gls{idp} and submits a hashed-and-salted conversation's safety number with that request.
The \gls{idp} will then issue a token that includes a signature on the safety number and one of the prover's digital identities.
At its core, this signature enables SOAP to provide social authentication and hashing-and-salting the safety number provides privacy.
The prover forwards the token to the verifier, whose messaging application verifies it cryptographically and displays the prover's identity if all checks pass.
In particular, this means that neither the verifier nor the prover must interact with cryptographic objects such as cryptographic keys or fingerprints thereof.
In practice, users must run SOAP once per \gls{idp} to authenticate themselves to one contact, and only need to rerun SOAP should their long-term key material change.

We propose to run SOAP with multiple \glspl{idp} (as shown in Fig.~\ref{fig:app-setup}), which substantially improves user security compared to Signal's and WhatsApp's default security.
To the user, these multiple runs of SOAP (for multiple \glspl{idp}) will appear as one, which will become clear when we explain our prototypes in Section~\ref{sec:proto}.
Recall that, by default, a user's account can be attacked under the following condition: compromise the application's key server or compromise the cellular provider while the registration lock is not enabled (Sec.~\ref{sec:bg}).
Now suppose the prover runs SOAP (Step~5, Fig.~\ref{fig:app-setup}), both with Microsoft and GitLab as the \gls{idp}.
Both protocol runs are independent and, when completed successfully, the verifier's app will display both the prover's Microsoft and GitLab identity.

It is now much harder to impersonate or \gls{mitm} the prover: the adversary must compromise both Microsoft \textit{and} GitLab \textit{and} either the prover's cellular provider \textit{or} the application's key server.
Critically, the compromise of the application's key server no longer suffices, and in contrast to the registration lock, users can authenticate their chat partners and need not rely on their chat partners activating the registration lock.

\subsection{Threat Model}
\label{sec:adversary}
SOAP's security properties hold against two kinds of adversaries: We establish social authentication against an active network adversary and privacy against a malicious \gls{idp}.
Whereas the social authentication-adversary can read, intercept, reorder, and replay all messages, the malicious \gls{idp} can do the same but only with messages sent to it directly.
For example, the malicious \gls{idp} cannot observe whether the prover forwards tokens to the verifier.
We restrict our analysis of SOAP's privacy property to a malicious \gls{idp} as we wish to show that adding \glspl{idp} to the messaging application ecosystem does not threaten user privacy.

Our threat model permits the compromise of the messaging application's key server, the leaking of OpenID Connect requests to \glspl{idp}, and the compromise of some of the \glspl{idp} integrated into the messaging application.
Notably, we make no assumptions on user-behavior other than that users do not leak their credentials.
Users may click any link sent to them, whenever an \gls{idp} asks them for consent, they may provide it, and whenever an \gls{idp} asks them to log in, they may do so, even if the adversary triggered that query.
We only limit our adversaries' capabilities in the following ways:

\begin{enumerate}[noitemsep]
  \item \label{itm:symbolic-model} Adversaries are bound by the security properties of the cryptographic primitives used and the TLS and Signal protocols (both Signal and WhatsApp use the Signal protocol).
  For example, adversaries can neither invert cryptographic hash functions nor eavesdrop on a TLS session.
  \item \label{itm:credentials-uncompromised} User credentials at \glspl{idp} are uncompromised.
  \item \label{itm:idp-uncompromised} Whenever a user authenticates via a given \gls{idp}, that \acrshort{idp}'s signing keys and TLS certificates are uncompromised.
  \item \label{itm:fwd-confidential} The messaging application and the user's web browser are uncompromised.
  In particular, the parameters of browser redirects to the messaging application remain confidential until they expire, and messaging application key material remains uncompromised.
\end{enumerate}

The necessity for Assumptions~\ref{itm:symbolic-model}-\ref{itm:idp-uncompromised} is self-evident.
Regarding Assumption~\ref{itm:fwd-confidential}, it should be clear that we must require the messaging application and the user's browser to be uncompromised.
We will discuss why we require the parameters of browser redirects to remain confidential later in Section~\ref{sec:discuss-threat-model} as this assumption requires a deeper understanding of SOAP's design.
Finally, we require that messaging application key material remains uncompromised because an adversary could otherwise launch a trivial impersonation attack.
They could run SOAP with a victim's safety number as a parameter and use an attacker-controlled account to log in.
The adversary could then inject that token into the conversation between the victim and one of the victim's contacts as they compromised the key material.
However, note that \begin{enumerate*}[label=(\roman*)]
  \item messaging secret keys are stored on-device only, i.e., a compromise of these keys likely comes with a compromise of the messaging application, and that
  \item the Signal protocol offers strong security guarantees against such key compromise, namely forward secrecy and post-compromise security \cite{InteractiveSignalVerification}.
\end{enumerate*}
SOAP was not designed to defend against key compromise, but rather against impersonation by associating malicious keys with pseudonyms such as in the Twilio incident.

\section{Protocol Design}
\label{sec:design}
In this section, we present SOAP's design in detail.
SOAP utilizes the OpenID Connect protocol \cite{OpenIdConnect,OIDCDiscovery} to facilitate adoption, which we explain first.

\subsection{OpenID Connect}
\label{apx:oidc}
\begin{figure}
  \centering
  \includegraphics[width=\columnwidth]{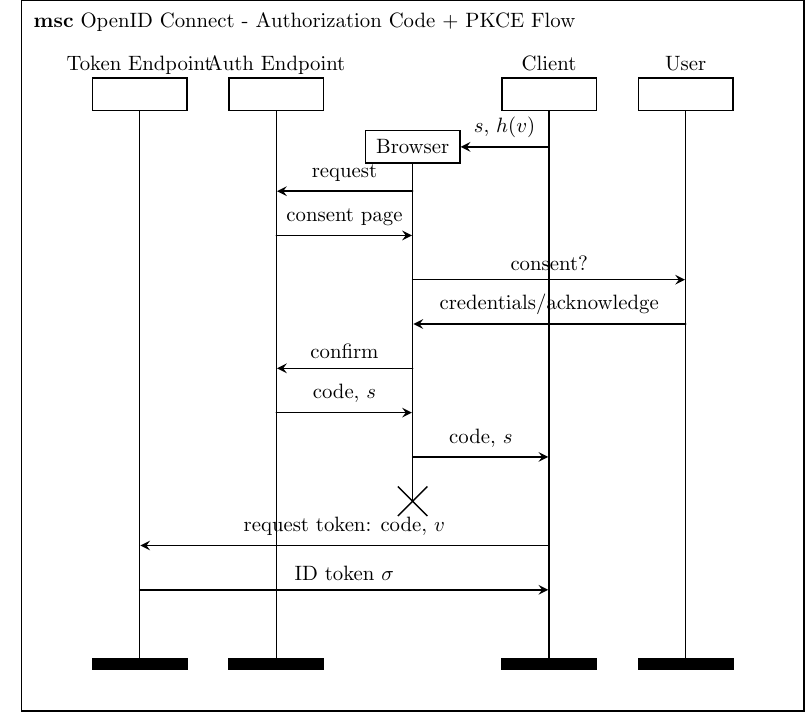}
  \caption{OpenID Connect authorization code flow with \acrshort{pkce}.
  $h(v)$ is the commitment to a random value $v$ as specified by \acrshort{pkce} and $s$ is the state parameter.}
  \label{fig:oidc-pkce}
\end{figure}

OpenID Connect \cite{OpenIdConnect,OIDCDiscovery} is an authentication protocol that itself builds on the OAuth 2.0 authorization protocol \cite{OAuthRFC}.
OpenID Connect is used to implement the well-known \textit{``Login with Google/Microsoft/Apple/\dots''} buttons.
OpenID Connect involves three parties: a \textit{user}, an \textit{\gls{idp}} managing the user's identity, and a \textit{relying party} seeking to authenticate the user.
The OpenID Connect protocol is executed by a \textit{client} operated by the relying party.
At the end of a successful protocol run, the client receives an \textit{ID token} through a browser redirect from the \gls{idp}.
The ID token is a cryptographically signed message, proving that the \gls{idp} authenticated the respective user, and which the client can use to identify the user.
Prior to issuing requests, relying parties must register at the respective \gls{idp}.
During registration, relying parties whitelist redirect URLs, and \glspl{idp} issue \textit{client IDs}.

OpenID Connect supports multiple \textit{flows}, which are protocol variants aiming at specific types of software clients.
We use the authorization code flow extended with the \gls{pkce} standard \cite{PKCE}, which is currently recommended as best practice for clients such as mobile applications \cite{OAuthSecBestPractices}.
The authorization code flow with \gls{pkce} implements a commitment reveal scheme, which we depict in Figure~\ref{fig:oidc-pkce} and works as follows.
The client first issues an authorization request by launching the device's browser at a specific URL called the \textit{authorization endpoint}.
The request encodes various parameters in the URL: a client ID, redirect URL, code challenge (the commitment, which is the hash of a random string), and optionally a state parameter and a nonce.
The state parameter and nonce protect against replay and \gls{csrf} attacks.

After receiving an authorization request, the \gls{idp} verifies the redirect URL as whitelisted for the given client ID, authenticates the user, and asks the user for consent.
Users usually authenticate by logging in, or through a session cookie already stored in their browser.
Depending on the user's history with the \gls{idp}, the user may not need to grant consent.

Once a user consents to logging in, the \gls{idp} forwards the browser to the redirect URL given in the request.
In the redirect URL, the \gls{idp} encodes an \textit{authorization code} and the state parameter sent earlier.
The client verifies that the state matches the state it previously issued, and exchanges the authorization code for an ID token.
The client does this by sending a POST request to the \acrshort{idp}['s] \textit{token endpoint}, including the authorization code and the \textit{code verifier}.
The code verifier opens the code challenge commitment sent earlier to the \gls{idp}.
This allows the \gls{idp} to determine that the ID token is requested from the same client that issued the initial request.

The ID token is encoded as a \gls{jws}, which is a signed object that maps keys to values.
Amongst other values, the object includes the issuer, the audience (identifying the client), the subject (the user who was authenticated), the nonce, and a validity period.
Usually, ID tokens are short-lived, with lifetimes typically ranging from two minutes to two hours.
According to the OpenID Connect specification \cite{OpenIdConnect}, ID tokens must only be accepted by the intended audience.
This prevents a service to which a user logged in from using the ID token with another service.

\begin{figure}
  \centering
  \includegraphics[width=\columnwidth]{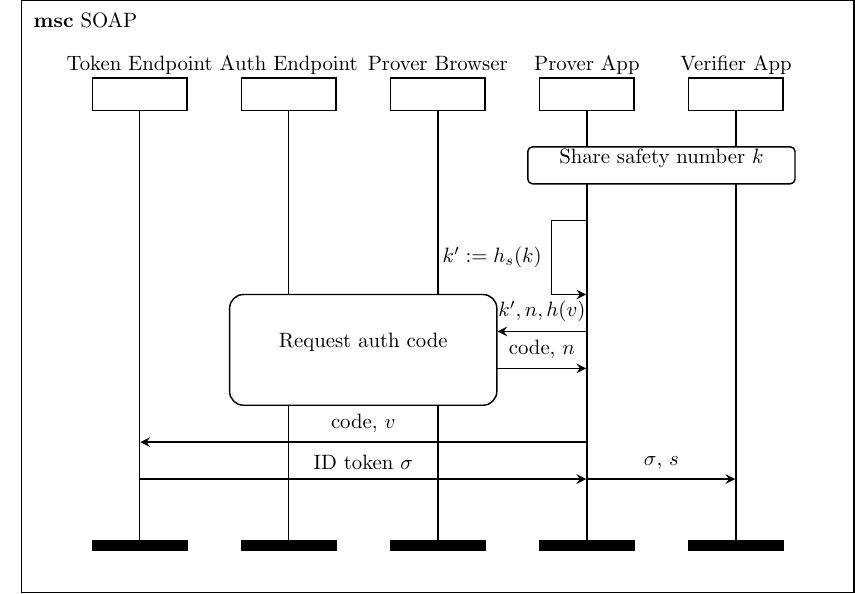}
  \caption{
    SOAP running between the prover, the \gls{idp}, and the verifier.
    Here, $h_s$ is a password hashing algorithm using a salt $s$ and $h$ is SHA-256 as specified by \gls{pkce}.
    The application randomly samples a code verifier $v$, a salt $s$, and nonce $n$.
    Finally, $\sigma$ is the OpenID Connect token, which is forwarded to the verifier and includes a signature on $h_s(k,s)$ and $n$.
  }
  \label{fig:core-protocol}
\end{figure}

\subsection{Protocol Description}

We next present SOAP, which provides an automated social authentication ceremony that requires no \gls{idp} adoption.\footnote{%
  For a technical description, see Appendix~\ref{apx:protocol}.
}
We suggest running SOAP with multiple \glspl{idp} (Section~\ref{sec:idea-and-props}), and that applications provide a user experience where these multiple runs appear as one.
Users are then only required to select \glspl{idp} and to provide consent.
We base SOAP on the OpenID Connect authorization code flow with \gls{pkce} \cite{OpenIdConnect}.
This makes its adoption straightforward, and it can be used immediately with any \gls{idp} that supports OpenID Connect.

The idea behind SOAP is to utilize the signed ID token to convince the verifier that the prover
\begin{enumerate*}[label=(\roman*)]
  \item authenticated to the \gls{idp} using the identity included in the token, and
  \item submitted their session's safety number when logging in.
\end{enumerate*}
The verifier must only check that the token was intended for them by comparing the safety number included in the token to the safety number of the session through which they received the token.
SOAP consists of three steps for the prover (request, validation, and forwarding) and one step for the verifier (validation).
Figure~\ref{fig:core-protocol} sketches our protocol.

To start a run of SOAP, the prover's messaging application prepares the request.
It generates two random values: a code verifier and a nonce $n$.
The application then uses a secure password-hashing algorithm to calculate a salted hash $h$ of the safety number.
This hash serves to blind the safety number to the \gls{idp}.
To defend against \glspl{csrf} attacks, the application stores the nonce, the salt, the \acrshort{idp}'s ID, and the code verifier as the most recently issued request.
Then, the application launches the authorization code flow, passing $n$ as the nonce and as the state, $h$, the code challenge, and a redirect URL.
This redirect URL must be distinct for each \gls{idp} and should use the HTTPS scheme (preferred) or a custom scheme.
The application then launches the system's browser with the request URL, which takes the prover to the consent/login page.

If the prover consents, the \gls{idp} redirects the browser back to the application.
The redirect passes the application a state parameter and an authorization code, which can be exchanged for an ID token.
Before the application uses the authorization code, it must verify that the state value it just received equals the nonce stored with the most recently issued request, and that the response originates from the expected \gls{idp}.

If both checks pass, the application uses the authorization code and stored code verifier to request the ID token from the \gls{idp}.
It verifies the token's signature and fields, e.g., that it correctly encodes the hashed and salted safety number.
Finally, the application clears its storage for the most recently issued request, and stores the nonce in a replay cache.
Recording nonces defends against reflection attacks; the description of our web-based prototype in Section~\ref{sec:web-proto} illustrates this threat.

The application forwards the ID token and the salt to the verifier and the verifier applies the same checks.
It also verifies that the safety number encoded in the ID token encodes the prover and verifier's keys, and that it did not request this token itself by looking up the stored nonces from runs where it was the prover.
If these checks pass, the verifier can obtain the sender's identity from the token.

While Figure~\ref{fig:core-protocol} may suggest that SOAP simply ``calls OpenID Connect,'' previous work \cite{OAuthSecBestPractices,WIMOauth2,WIMOpenID} highlighted many subtleties in implementing an OpenID Connect-based protocol securely.
Hence, we next present our formal proof that SOAP indeed provides social authentication.

\section{Security Analysis}
\label{sec:security}
SOAP is designed to implement social authentication and to protect user privacy, as defined in Section~\ref{sec:prop}.
In the following, we prove its security using the protocol-verifier Tamarin \cite{Tamarin}.
Tamarin has been used to verify many critical protocols, such as 5G-AKA, TLS 1.3, and the credit card protocol EMV \cite{TamarinEMV,Tamarin5GAKA,TamarinTLS}.
We first introduce Tamarin, and then present the formal proofs of SOAP's social authentication and privacy property.

\subsection{The Tamarin Prover}
\label{sec:tamarin}

Tamarin \cite{Tamarin} is an infinite-state protocol verifier that supports both fully automated and guided proof construction in the symbolic model.
Tamarin models have three parts: rules modeling protocol steps, an equational theory modeling cryptographic primitives, and lemmas modeling protocol properties.

Rules have the form \lstinline{l --lbl-> r}, where \lstinline{l} reads from the current state, \lstinline{r} updates the new state, and \lstinline{lbl} labels this transition for reference in properties.
Typically, \lstinline{l} encodes reading incoming messages and parts of the participant's state and \lstinline{r} encodes sending messages and updating the participant's state.
For example, the rule below models that a participant receives a message, looks up a key, and sends the message symmetrically encrypted under the respective key.
The label adds \lstinline{Enc(m)} to the protocol trace for reference in properties.

\begin{lstlisting}[gobble=2,basicstyle=\ttfamily\small]
  [In(m),!K(k)] --[Enc(m)]-> [Out(senc(m,k))]
\end{lstlisting}

Equations model cryptographic primitives.
For example, the equation \lstinline{sdec(senc(m, k), k) = m} models that the symmetric decryption of a ciphertext using the correct key yields the respective plain text.
Equational theories allow one to formalize a strong network adversary that can derive new knowledge from previously sent messages.
For example, if a participant sent both a symmetrically encrypted message and the respective key, the aforementioned equation would allow the adversary to learn that message's content.

Tamarin supports two types of properties: universally quantified and existentially quantified trace properties.
Tamarin verifies both kinds of properties similarly.
Universally quantified properties are first negated, whereas existentially properties are left as is.
Then, Tamarin tries to construct a trace that satisfies the resulting property using backwards constraint solving.
For universally quantified properties, Tamarin hence tries to construct a counterexample, and for existentially quantified properties, Tamarin tries to construct a positive example.

\subsection{Social Authentication}
\label{sec:authentication}

We next provide high-level intuition on why SOAP provides social authentication and then describe our formal proof.

\subsubsection{Informal Analysis}
\label{sec:sec-arg}

We previously defined social authentication as an implication: ``If the verifier associates an account $A$ with the public key $PK$ and received a message from each of these pseudonyms, then these messages were sent by the same party.''

There are two ways to violate social authentication.
Either, there is no send event for one of the messages or the send events have different senders.
In our formal model, we modelled the message-exchange channels as authentic (i.e., given a receive event, there will be a send event) and thus focus on the latter case.
Given SOAP's design, the attacker can achieve this by either making the prover send a malicious token through the messaging application, linking an attacker-chosen account to the prover's messaging channel (identity substitution), or making the prover authenticate in an attacker-controlled OpenID Connect flow, linking the prover's account to an attacker-chosen channel (impersonation).

Let us first focus on impersonation attacks, and presume Eve attempts to attack Alice.
This means that Eve has a messaging session with Alice and wants to convince Alice that she, Eve, controls one or more of Bob's accounts at \glspl{idp}.
To achieve this, Eve must send a token that includes a reference to Bob's account and the safety number of Eve's and Alice's public keys.
As we assume that Bob's account is uncompromised (Assumption~\ref{itm:credentials-uncompromised}, Sec.~\ref{sec:adversary}), Eve must craft a malicious OpenID Connect request and forward it to Bob.
This is straightforward as it requires nothing more than convincing Bob to click on a malicious link encoding such a request.
However, Eve cannot obtain a token from this: Bob's application will discard the authorization response if it did not issue the request itself.
Moreover, if it did issue the request itself, it would include one of Bob's safety number and not Alice and Eve's.

Similarly, Eve cannot launch an identity substitution attack.
Eve would need to run SOAP herself, using a victim's safety number as parameter, log in with her own account, and then make the victim's application accept the resulting browser forward.
However, the application would again discard that forward as it did not issue the corresponding request.

This argument spells out the main idea behind SOAP's security.
In reality, the security of protocols based on OAuth 2.0 and OpenID Connect is much more subtle.
Attackers can in general access authorization code responses through other means than capturing the redirects.
For example, \cite{WIMOauth2} first described the \gls{idp}-mixup attack, in which applications leak an authorization code by sending it to the wrong \gls{idp}.
Therefore, we formally evaluate SOAP's security next.

\begin{figure}
  \centering
  \includegraphics[width=.95\columnwidth]{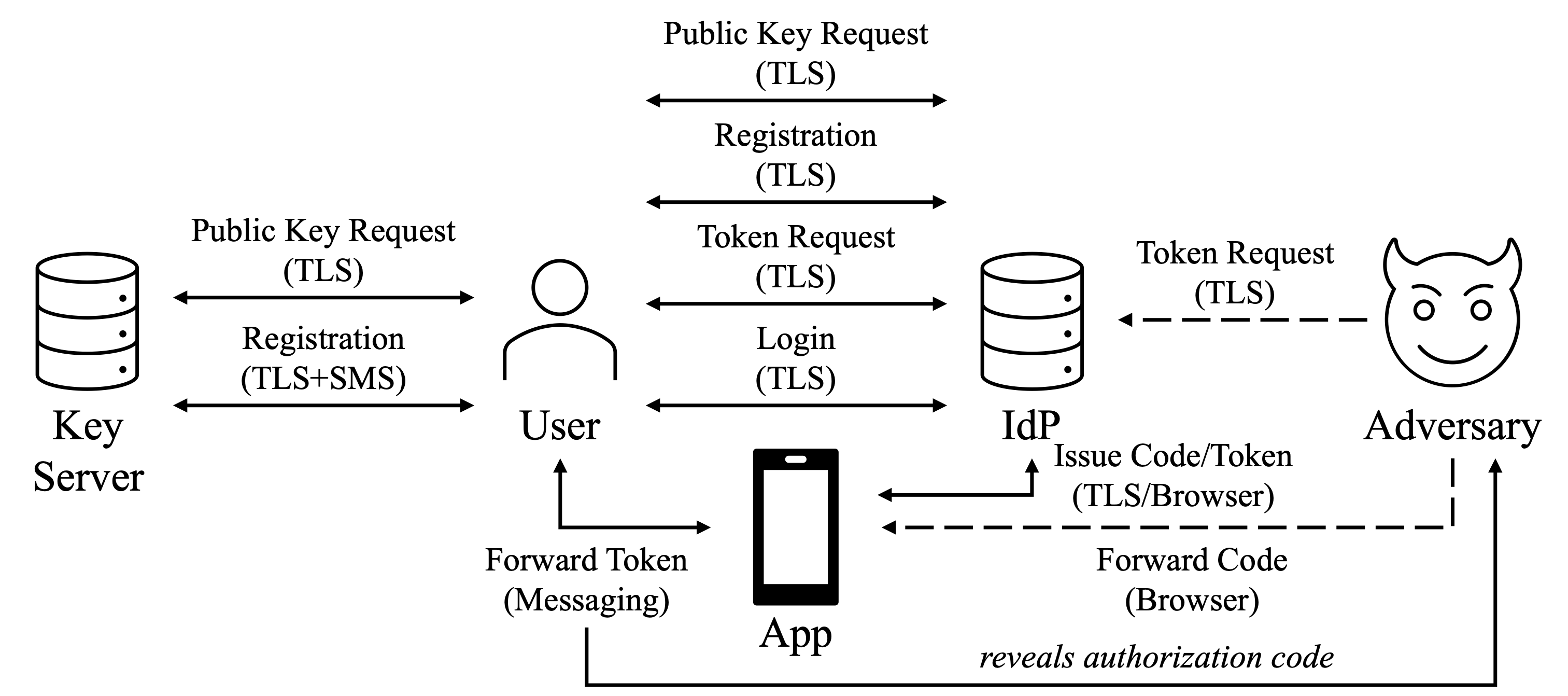}
  \caption{Sketch of our formal model of SOAP.
  Arrows indicate message exchanges, denoted with the respective channels used.
  Dashed arrows indicate that the adversary can initiate the respective request on the user's behalf.
  We omit the \gls{idp}-controlled bulletin board and the messaging channel.}
  \label{fig:model-sketch}
\end{figure}

\lstset{
  backgroundcolor=\color{white},
  frame=tb,
  keywords={All,Ex,rule},
  keywordstyle=\color{blue}\bfseries,
  literate=%
    {\&}{{{\color{blue}\&}}}1
    {|}{{{\color{blue}|}}}1
    {==>}{{{\color{blue}==>}}}3,
  numbersep=5pt,
  numberstyle=\tiny\color{gray},
  rulecolor=\color{black},
}
\begin{figure*}
  \centering
  \begin{lstlisting}[gobble=4,numbers=left,escapechar=*]
    All v sendKey rcvKey m1 idp acc m2 #t #r1 #r2.
        ( Correspond(v, sendKey, idp, acc) @ #t *\label{ln:associate}*
        & ReceiveMessaging(sendKey, rcvKey, m1) @ #r1 *\label{ln:rcv-msg}*
        & ReceiveIdP(idp, acc, m2) @ #r2)
    ==> ( (Ex s #x1 #x2.
          ( SendMessaging(sendKey, rcvKey, m1) @ #x1 & Sender(s) @ #x1 & #x1 < #r1 *\label{ln:sent-msg}*
          & SendIdP(idp, acc, m2) @ #x2 & Sender(s) @ #x2 & #x2 < #r2)) *\label{ln:sent}*
        | (Ex p #x. CompromisedAccount(p, idp, acc) @ #x) *\label{ln:acc-compr}*
        | (Ex #x. CompromisedIdP(idp) @ #x) *\label{ln:idp-compr-1}*
        | (Ex #x. CompromisedDomain(idp) @ #x) *\label{ln:idp-compr-2}*
        | (Ex app redirectURL #x #y #z. *\label{ln:app-compr}*
              IsMessagingApp(app) @ #x
            & IsRedirectURL(idp, app, redirectURL) @ #y
            & CompromisedDomain(redirectURL) @ #z)
        | (Ex p #x. CompromisedMessaging(p, sendKey) @ #x)) *\label{ln:key-compr}*
  \end{lstlisting}
  \caption{Formalization of social authentication (Sec.~\ref{sec:prop}), also encoding the threat model (Sec.~\ref{sec:adversary}).
  Note that the two \lstinline{Sender} facts are bound to the respective \lstinline{SendMessaging} and \lstinline{SendIdP} events, as they occur at the same time points (\lstinline{x1} and \lstinline{x2}).
  \label{fig:prop-formal}}
\end{figure*}

\subsubsection{Formal Proofs}
\label{sec:analysis}

Our formal model comprehensively captures SOAP and its heterogeneous environment.
Beyond the messaging application and TLS, we modelled all security-critical aspects and auxiliary protocols such as public key requests and distribution.
Figure~\ref{fig:model-sketch} depicts a sketch of what we modelled.
Our model permits arbitrarily many participants to communicate with each other in arbitrarily many parallel protocol sessions.
The adversary can corrupt any party in a fine-grained manner, e.g., a user's account could be compromised independently of their messaging long term keys, only constrained by our security assumptions from Section~\ref{sec:adversary}.
For example, \gls{idp} corruption is possible for every \gls{idp} except the one with which the prover intends to authenticate themselves.

More specifically, our model includes channels for SMS, TLS, browser redirects, and messaging applications with different security properties and distinct keys.
We also modelled communication associated with \gls{idp}-controlled pseudonyms (usernames) as a bulletin board where users can post messages associated with their pseudonym publicly after the \gls{idp} authenticates them using a password.
We modelled SMSes as insecure and the messaging application as secure (confidential and authentic).
We modelled TLS as a secure channel without client authentication, i.e., the adversary can always initiate new sessions with servers.

The adversary can compromise any TLS server, which allows it to read client queries and respond to them.
TLS queries can have one of two methods, GET and POST, whereby GET requests can be initiated by the adversary on a user's behalf, modelling that the adversary can trick users into clicking any link.
In practice, this allows the adversary to launch the OpenID Connect protocol at various points (e.g., initial request and code forwarding) for non-compromised clients.
Browser redirects are modelled as GET requests using an existing session to connect to a new server, initiated by the previous server.
This allows us to model, e.g., the redirect to a mobile application by modelling that application as a server.

We modelled the messaging provider, messaging application, end users, and \glspl{idp} as different parties.
Our model includes SOAP itself, messaging application registration (including SMS \gls{otp} verification), \gls{idp} account registration, as well as messaging key server and \gls{idp} public key requests and responses.
Moreover, we fully modelled OpenID Connect and we make no assumptions on this protocol's security.

Within this model, we prove that SOAP implements social authentication.
Figure~\ref{fig:prop-formal} shows our Tamarin specification formalizing social authentication as a trace property.
It has three parts.
Lines~\ref{ln:associate}-\ref{ln:sent} formalize social authentication: If the verifier associates two pseudonyms with each other (\lstinline{Correspond}), then all messages received from those two pseudonyms (\lstinline{ReceiveMessaging}/\lstinline{IdP}) originate from the same sender.
We formalize the latter by showing that there exist two send events (\lstinline{SendMessaging}/\lstinline{IdP}) for which the sender (\lstinline{Sender}) is the same party \lstinline{s}.

Lines~\ref{ln:acc-compr}-\ref{ln:key-compr} formalize Assumptions~\ref{itm:credentials-uncompromised}-\ref{itm:fwd-confidential} from our threat model (Sec.~\ref{sec:adversary}).
Assumption~\ref{itm:symbolic-model} is covered implicitly as Tamarin operates in the symbolic model.
There is just one subtle difference from social authentication as presented earlier.
Namely, since the messaging channel is not just sender-authenticated, but also receiver-authenticated, we can include the recipient's pseudonym \lstinline{rcvKey} in the receive and send event in lines~\ref{ln:rcv-msg} and \ref{ln:sent-msg}.
This means that our formalization of social authentication is stronger than sender correspondence.

The size and complexity of our model put its security properties out of reach for fully automated verification.
For example, modelling that browser redirects remain confidential \textit{until they expire} (Assumption~\ref{itm:fwd-confidential}) proved to be challenging.
In part, we modelled this assumption by revealing authorization codes to the adversary after receiving the respective ID token.
This led to infinite looping in Tamarin's proof construction, which we avoided by proving an inductive, auxiliary lemma showing that authorization codes can only be used once.
In total, we verified nine auxiliary lemmas and programmed custom proof heuristics to aid Tamarin's proof construction.

\subsubsection{Discussion of Threat Model}
\label{sec:discuss-threat-model}

With our formal analysis of the authentication property completed, we briefly return to Assumption~\ref{itm:fwd-confidential} of our threat model, where we require that the parameters of browser redirects to the messaging application remain confidential.
Without this assumption, the adversary could easily obtain an identity token that binds an attacker-chosen safety number to a victim's account.
To achieve this, they would only need to trick their victim into clicking a SOAP-request link that includes a malicious safety number as parameter.
As soon as the victim logs in and consents (which they will do under our liberal threat model), the adversary could learn the authorization code from the redirect URL and request the ID token themselves.

Users can theoretically protect themselves from this attack by only granting consent to requests they initiated themselves.
However,
\begin{enumerate*}[label=(\roman*)]
  \item we find it unrealistic to assume users are resistant to social engineering, and
  \item we experienced during our prototype development that some \glspl{idp} immediately acknowledge requests without user involvement whenever the user was already logged in and had previously granted consent.
  In this case, users could be attacked easily, e.g., with malicious URLs obfuscated with a URL shortener.
\end{enumerate*}

In practice, though, capturing redirects requires the adversary to have access to a user's browsing history while an attack is launched.
This requires the compromise of the user's browser, or the installation of a malicious application handler on the user's device.
In case of a compromised browser, it is nigh impossible to protect the user's account credentials at the same time.
To address malicious application handlers, we recommend that the application should use HTTPS redirect URLs as described in Section~\ref{sec:design}.
With HTTPS URLs, application developers can utilize the security features provided by modern operating systems to ensure that authorization codes do not leak.
For example, both Windows and Android support applications to handle HTTPS URLs, but only as long as these applications have been delegated to do so by the respective URL \cite{AndroidLinks,WindowsLinks}.
This way, application developers can rely on the security features provided by the Web PKI to protect authorization responses.
Without HTTPS URLs, an attacker would still need to install a malicious application handler and in turn a malicious application on their victim's device to capture redirects to custom schemes.

\subsection{Privacy}
\label{sec:privacy}

SOAP protects users' privacy against the \glspl{idp} in that it only reveals that a prover is using SOAP, which messaging application is being used, and at what times SOAP is used.
We formally proved SOAP's privacy as an observational equivalence property using Tamarin.
Implementing our threat model (Sec.~\ref{sec:adversary}), we proved privacy in a simplified model (compared to the model presented in the previous section) that only includes communication between users and the \gls{idp}.
We modeled the malicious \gls{idp} as the adversary and consequently replaced TLS with an insecure channel.
Our observational equivalence property shows that \glspl{idp} cannot distinguish protocol runs where a user submits the correct salted-and-hashed prover/verifier safety number from runs where the user submits a different safety number.
In the symbolic model, our privacy property is straightforward as we will lay out next.

SOAP only includes two requests to the \gls{idp}['s] servers, and these requests include the following parameters: the messaging application ID with the \gls{idp}, a redirect URL, a nonce, the code challenge (the hashed code verifier), the code verifier, the authorization code, and the salted-and-hashed safety number.
The messaging application's ID and redirect URL reveal the messaging application and that the prover is using SOAP.
The nonce, the code challenge, and the code verifier are randomly generated values that change with every request, and hence, leak nothing about the prover.
The authorization code is issued by the \gls{idp} and therefore allows the \gls{idp} to connect the initial authorization request with the token request.
Finally, the salted-and-hashed safety number leaks nothing about the prover under the assumption that the \gls{idp} cannot break cryptographic primitives such as salt-and-hashing, and because the salt is only shared with the verifier.

In theory, a verifier could share the salt with an \gls{idp} to reveal that the prover intended to communicate with the verifier.
However, a verifier could leak this information even without SOAP.
The \gls{idp} could also attempt to correlate public key requests with issued tokens.
Fetching public keys, however, is not part of SOAP itself as applications will likely cache public keys.
Additionally, the OpenID Connect specification \cite{OIDCDiscovery} requires that \glspl{idp} distribute their public keys publicly and application-independently via HTTPS.
For both of these reasons, we deem such a correlation to be practically infeasible.
To summarize: SOAP prevents the surveillance by an \gls{idp} of its user's contact graphs in messaging applications.

\section{Implementation}
\label{sec:proto}
We implemented our proposal in two prototypes: as a stand-alone web application\footnote{%
  \label{ftnt:web-proto}
  The prototype is hosted at \url{https://soap-proto.net}.
} and as a fork of the Signal open-source Android application.
Both prototypes support GitLab and Microsoft as \glspl{idp}.
The stand-alone version does not require messaging application adoption but requires more user interaction.
In its current design, it can be used to associate arbitrary statements to a user's account.

Our prototypes demonstrate that social authentication can be realized practically, with provable security guarantees, and without OpenID Connect-\gls{idp} adoption.
Our web-based prototype shows that social authentication can be implemented even without messaging application adoption.
Library support for OpenID Connect is abundant and, hence, adoption is straightforward.
One of the authors could implement an initial prototype within a day.
With just a few clicks and in a couple of seconds, users can verifiably share their identity.

\subsection{Web-based Prototype}
\label{sec:web-proto}

\begin{figure}
  \centering
  \begin{subfigure}[t]{.48\columnwidth}
    \centering
    \fbox{\includegraphics[width=\textwidth]{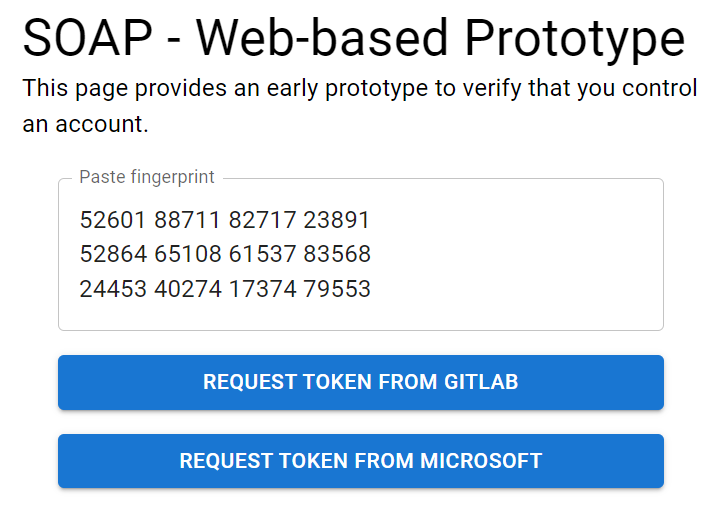}}
    \caption{Prover pastes safety number into the application and selects one of two \glspl{idp}.}
    \label{fig:js-proto-init}
  \end{subfigure}
  \hfill
  \begin{subfigure}[t]{.48\columnwidth}
    \centering
    \fbox{\includegraphics[width=\textwidth]{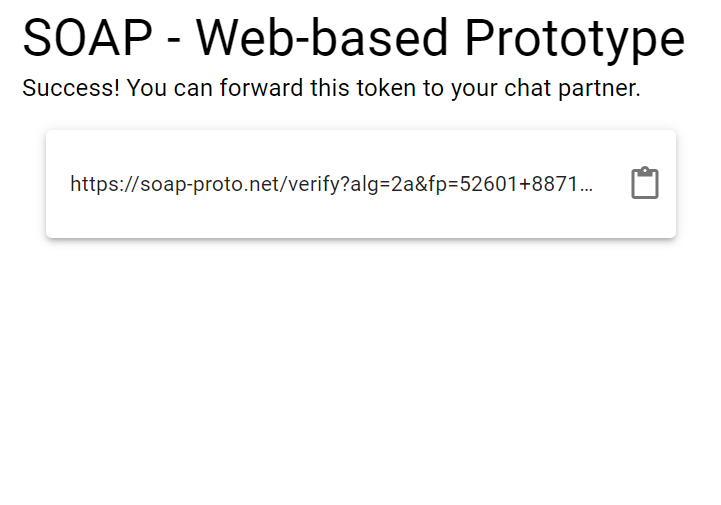}}
    \caption{Prover receives ID token and is presented a URL to forward to their chat partner.}
    \label{fig:js-proto-fwd}
  \end{subfigure}

  \begin{subfigure}[t]{.48\columnwidth}
    \centering
    \fbox{\includegraphics[width=\textwidth]{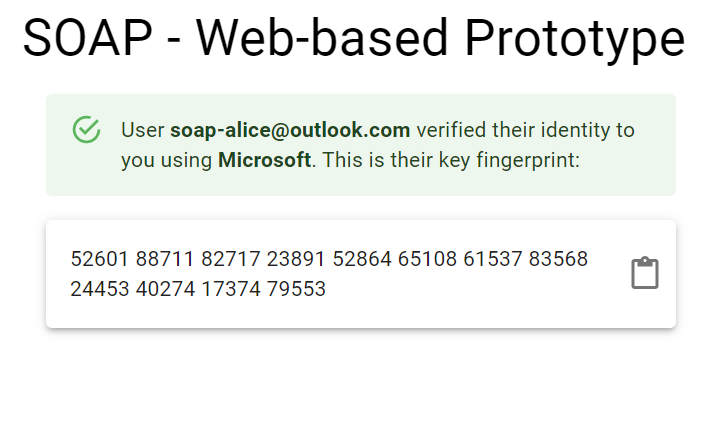}}
    \caption{Verifier sees the prover's identity and safety number when clicking the link.}
    \label{fig:js-proto-recv}
  \end{subfigure}
  \hfill
  \begin{subfigure}[t]{.48\columnwidth}
    \centering
    \fbox{\includegraphics[width=\textwidth]{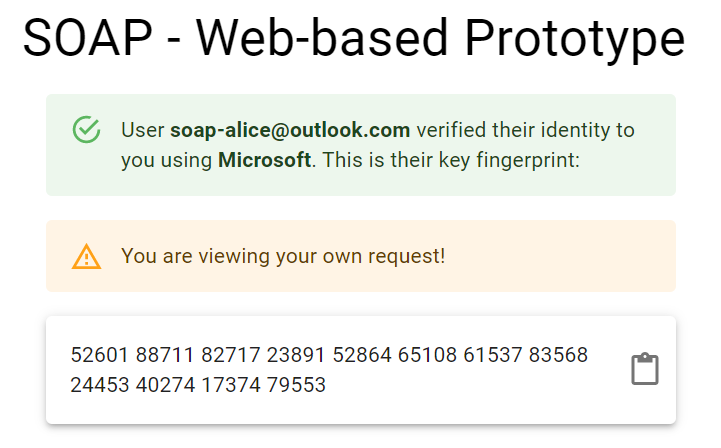}}
    \caption{Prover sees a warning when clicking a link generated within the same browser.}
    \label{fig:js-proto-warn}
  \end{subfigure}
  \caption{Screenshots depicting the web application prototype.
  The prover must initiate this flow for each \gls{idp}.}
  \label{fig:js-proto}
\end{figure}

Figure~\ref{fig:js-proto} depicts the interaction with our web-based prototype.
When started, the application renders a text input and a list of \glspl{idp} to select from (Fig.~\ref{fig:js-proto-init}).
After the prover selects an \gls{idp}, the application assumes that the input contains the safety number to authenticate, and initiates SOAP.
After they complete the OpenID Connect flow (see Sec.~\ref{apx:oidc}), the prover is forwarded to our application, which requests the ID token (Fig.~\ref{fig:js-proto-fwd}).
The web application verifies the ID token, and, if all checks pass, displays a success message.
The prover can then copy a link to send to their chat partners.

This link encodes the ID token and everything needed to verify it.
When the verifier clicks the link, the application verifies the token, and if all checks pass, displays the prover's identity, the \gls{idp}, and the safety number.
The verifier can copy this safety number to their clipboard.
The Signal Android application offers its users to compare the safety number with the clipboard, giving the verifier an easy way to check the safety number (Sec.~\ref{sec:bg}).

Note that the application will render a success message to anyone clicking a link containing an ID token.
Only the user can determine whether the safety numbers match.
This allows for social engineering attacks whenever users click links they earlier forwarded themselves.
To mitigate this threat, the web application warns users that they issued this request themselves whenever they click a link that was issued within the same browser.
Figure~\ref{fig:js-proto-warn} shows the view after clicking on a URL that encodes an ID token.

\subsection{Signal Prototype}

\begin{figure}
  \begin{subfigure}[t]{.3\columnwidth}
    \centering
    \fbox{\includegraphics[width=\textwidth]{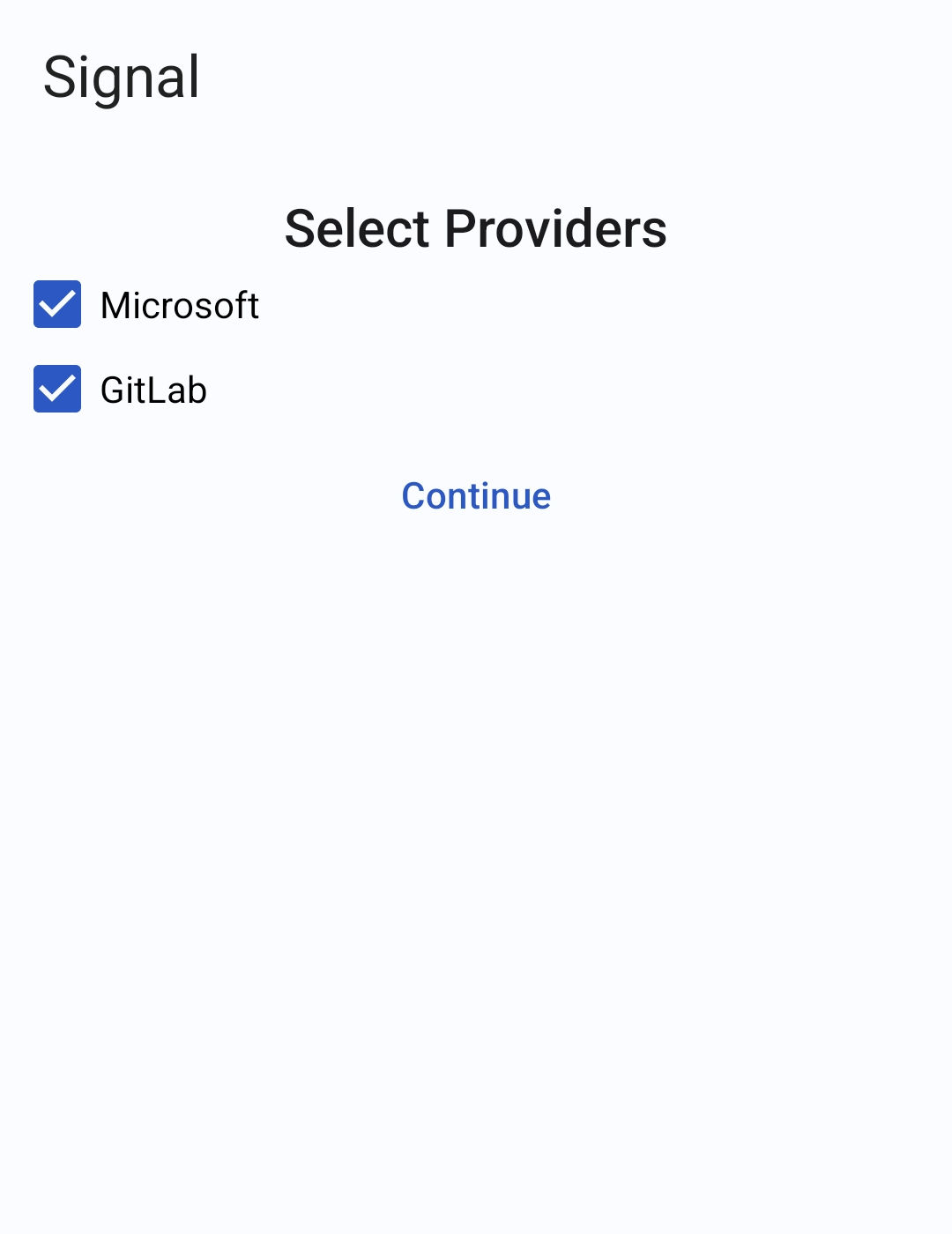}}
    \caption{Prover selects the \glspl{idp} to authenticate with.}
    \label{fig:signal-proto-select}
  \end{subfigure}
  \hspace{5pt}
  \begin{subfigure}[t]{.3\columnwidth}
    \centering
    \fbox{\includegraphics[width=\textwidth]{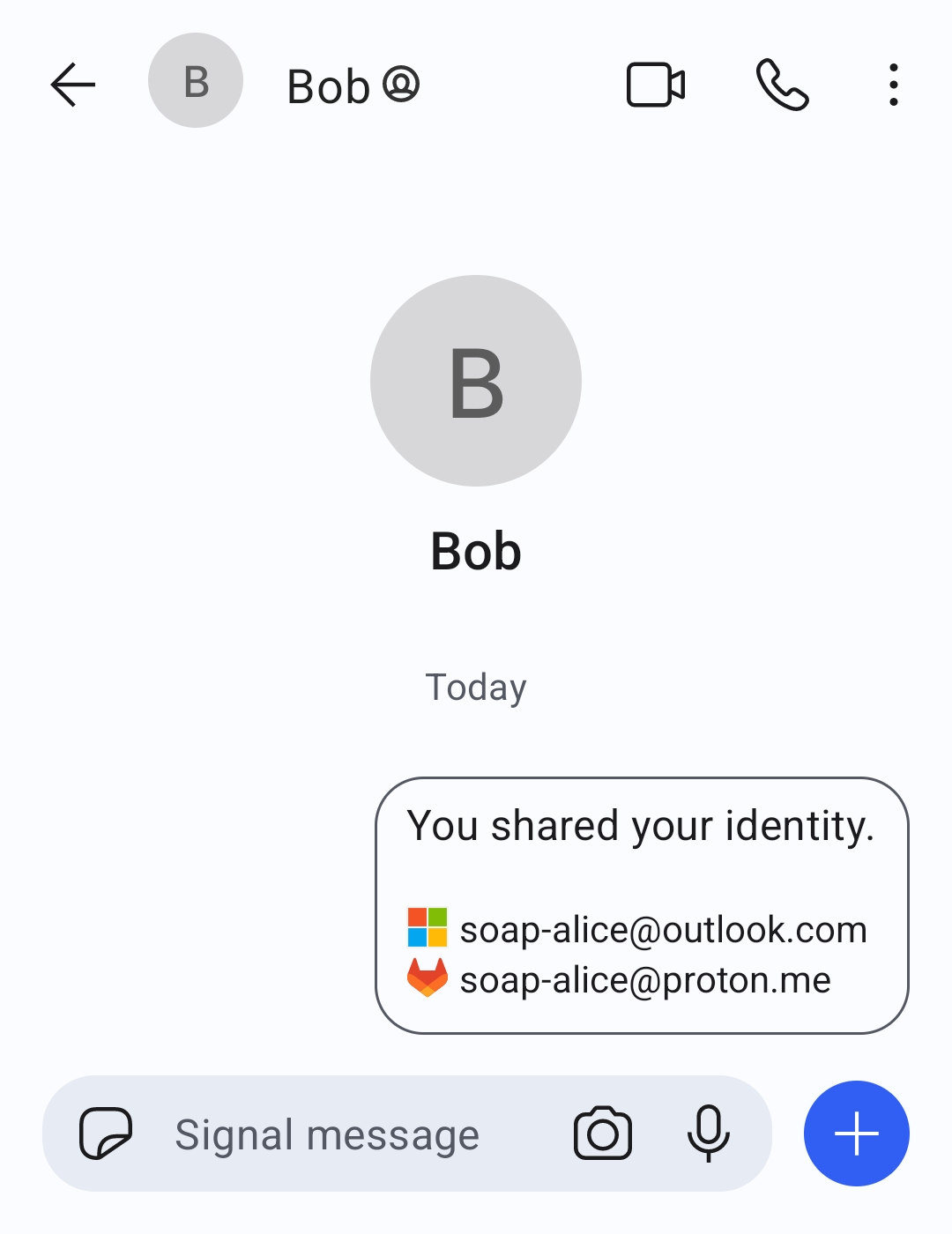}}
    \caption{Prover sees that they shared their identities.}
    \label{fig:signal-proto-success-sender}
  \end{subfigure}
  \hspace{5pt}
  \begin{subfigure}[t]{.3\columnwidth}
    \centering
    \fbox{\includegraphics[width=\textwidth]{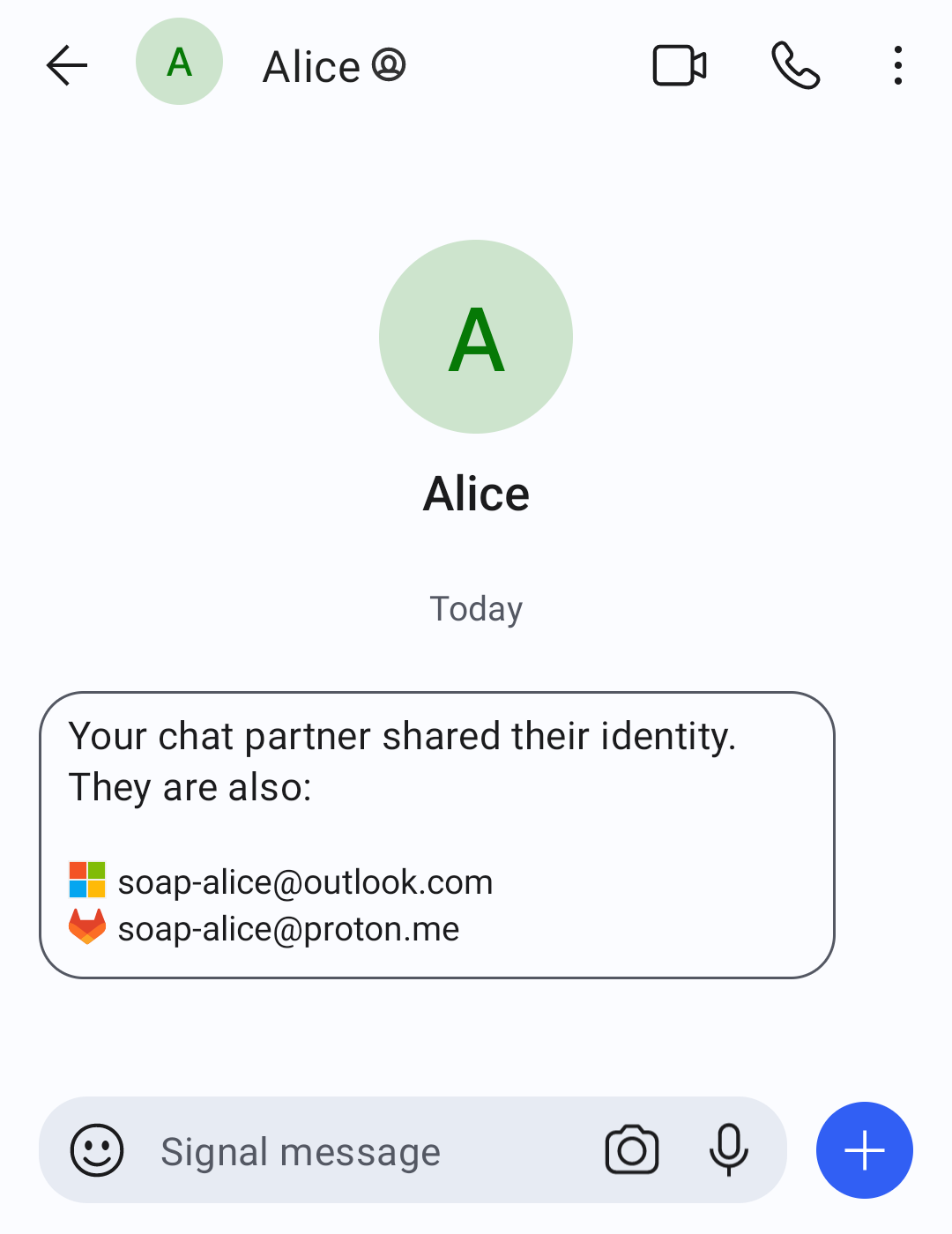}}
    \caption{Verifier sees the prover's identities.}
    \label{fig:signal-proto-success-receiver}
  \end{subfigure}
  \caption{Screenshots depicting the Signal prototype.
  The interaction with our prototype is depicted from left to right.}
  \label{fig:signal-proto}
\end{figure}

Our Signal prototype, depicted in Figure~\ref{fig:signal-proto}, provides a more streamlined experience compared to the web-based prototype.\footnote{%
  A video demo of the Signal prototype can be viewed at \url{https://youtu.be/Ip_RAF8PRrM}.
}
In particular, the Signal prototype requires significantly less interaction from the prover and no interaction from the verifier.
Users need not actively examine and insert safety numbers, and flows need not be initiated for each \gls{idp}.

When a user wishes to authenticate themselves to their chat partner (becoming a prover), they must first select the new ``Authenticate'' option within the attachment menu.
Next, the prover selects all the \glspl{idp} they wish to authenticate themselves with (Fig.~\ref{fig:signal-proto-select}).
When they press continue, they will run SOAP for each of the \glspl{idp}.
The application verifies all tokens and displays a distinctly styled message to both the prover and verifier that shows the shared identities (Fig.~\ref{fig:signal-proto-success-sender}, \ref{fig:signal-proto-success-receiver}).

Our Signal prototype demonstrates that SOAP is a practical design that requires little user interaction.
While it is adequate for a proof-of-concept, we suggest further enhancements before it is deployed in production.
First, previously performed social authentications should be recorded as such in the application.
The application could display a contact's identities within the chat header, rather than mark them as ``Verified.''
This would additionally highlight that our proposal is intended to augment in-person safety number comparison, not supersede it.
Second, the messaging application could compare the identities provided through SOAP with identities linked to the contact's address book entry on the smartphone to automatically combat impersonation attacks.
Messaging applications are usually granted access to address books anyways, providing a straightforward means to automate social authentication further.

\subsection{Development}

The web-based prototype was written as a single-page JavaScript application in React \cite{React}, i.e., all code runs in the user's browser.
The prototype consists of roughly 500 lines of code and the first version was developed within a day by one of this paper's authors.
In contrast to the web-based prototype, the Signal prototype's development was more involved and required around three person weeks.
Understanding the meagerly documented Signal Android codebase demanded most of the time.
We changed 21 files and added around 1000 lines of Java and Kotlin code for the application logic, and also changed 17 files with around 200 additional lines of code for configuration changes, like layout and localization.

During prototype development, we noticed that not all \glspl{idp} support OpenID Connect ideally for our use case.
Microsoft, for example, does not support HTTPS redirect URLs for Android applications.
We tried working around this restriction by registering our Android application as a Single-Page Application, which permitted us to configure an HTTPS redirect URL.
However, requesting the ID token failed as cross-origin request headers were missing.

GitLab supports HTTPS redirects and does not distinguish the types of applications upon registration.
However, GitLab does not ask for consent again after the user consented once to log in.
This causes HTTPS redirects to Android applications to fail.
A Chrome policy requires user interaction in order to redirect users to Android applications through HTTPS URLs \cite{AndroidIntentsChrome}.
This left us with using custom schemes, e.g., \verb|auth://|, in redirect URLs to support GitLab as an \gls{idp}.

Finally, Google neither allows one to configure a redirect URL nor lists a redirect URL when registering Android applications as an OpenID Connect client.
In this way, Google hides their OpenID Connect API.
We suspect this practice is intended to force developers to implement ``Sign-In with Google'' using Google's SDK \cite{GoogleSignInSDKAndroid}.
This SDK need not be configured with a redirect URL to request ID tokens and does not allow one to specify a nonce.

These findings suggest that while SOAP does not require adapting the OpenID Connect specification, explicating our use case in the OpenID Connect specification could still benefit users and developers.
Currently, users will only consent to log in, and developers have to use the nonce field outside its specified intent.
If OpenID Connect were to recognize user-submitted claims as a request parameter, users could grant consent to show that they control the given account and developers could use APIs supporting user-submitted claims.

\subsection{Performance}

The performance of both prototypes is mainly constrained by the page load times of OpenID Connect authorization endpoints, consent screens, and HTTP redirects.
We timed the prototypes with a user that was logged in and had authorized our application already.
In the web-based prototype, running SOAP for a single \gls{idp} was nearly instantaneous (<1 second), and in the Signal prototype, running SOAP for two \glspl{idp} consistently took less than 10 seconds.

As we described in Section~\ref{sec:design}, SOAP requires local, persistent storage to protect against replay and \gls{csrf} attacks.
SOAP stores the latest issued request, which only requires a small, constant amount of space, and the nonces generated by the application.
The number of nonces stored is limited by the number of SOAP sessions initiated by the user, and the nonces can be discarded after a token expires, which in our experience happens within two hours.

\section{Related Work}
\label{sec:rw}
\subsection{Key Authentication}

In this section, we examine three research strands aimed at improving key authentication:
key transparency, authentication ceremonies within messaging applications, and social authentication.

\paragraph{Key Transparency}
CONIKS, SEEMless, Parakeet, and KTACA \cite{Coniks,SEEMless,Parakeet,AnonymousKeyTransparency} implement \textit{key transparency} as an alternative to key authentication.
In key transparency, providers commit to a publicly auditable key directory, and users (or rather their messaging applications) both fetch their contacts' public keys from this directory and check that the public key associated with their pseudonym in the directory matches their actual public key.
Key transparency aims to obviate the need for key authentication as other users can assume that their peers would have taken action should their application detect a malicious public key associated with their pseudonym.
Recently, Meta announced that WhatsApp will deploy key transparency services and published an open-source key transparency service based on SEEMless and Parakeet \cite{WhatsAppKeyTransparency,WhatsAppKeyTransparencyCode}.

In contrast to SOAP, key transparency has the upside that it requires no user-interaction when there is no compromise, but this comes at the cost of a significant engineering effort and can require recruiting external auditors.
Moreover, key transparency only aims to protect against malicious providers and not outside attackers, such as in the Twilio incident (Sec.~\ref{sec:bg}), and key transparency cannot prevent compromise but only make it detectable.
Finally, key transparency provides no notion of authentication, which is desirable in its own right, e.g., to verify that one chats with the person controlling a given account on a different platform.
We see SOAP as complementing key transparency and not rivaling it.
Moreover, SOAP is far less complex than key transparency and, thus, it is simpler to deploy SOAP, especially for smaller organizations.

\paragraph{Authentication Ceremonies for Messaging Applications}
We already extensively compared SOAP to numerous manual authentication ceremonies in Section~\ref{sec:bg}.
Other researchers additionally investigated how one could improve authentication ceremonies to increase success rates.
\cite{ActionNeeded} proposed UI changes, and \cite{UserCentricDesignSignal} proposed new ceremonies altogether, but they did not consider whether those ceremonies were secure.
Both \cite{UserCentricDesignSignal,ActionNeeded} focus on manual verification, where users compare two pieces of information.
In contrast, SOAP lifts this burden from the user and instead asks them to authenticate a set of identities.

\paragraph{Social Authentication}
The idea of authenticating users using their profiles at \glspl{idp} was pioneered by Keybase \cite{Keybase}.
There, users can bind their social media accounts, e.g., at Twitter, to their Keybase account using so-called \textit{proofs} \cite{KeybaseDocs}.
Users do this by posting a signed message on a social media platform.
Other users can verify that a user linked their accounts by checking that the signature was generated using the key associated with the Keybase account, and posted by the claimed social media account.

\cite{SocialAuthentication} coined the term ``social authentication,'' proposed its application to Signal, and conducted a user study.
The authors found that users regard social authentication as understandable, easy to use, working asynchronously and remotely (in contrast to, e.g., in-person verification), and that it enhances their security when using multiple providers.
However, users gave social authentication a lower trust score than in-person verification, partially stemming from their limited understanding of the mechanism.
For example, users feared that a compromise of their social media accounts could lead to a compromise of their Signal account, and they distrusted social media providers in general.
Users also mentioned the risk of social engineering with fake accounts.

The authors of \cite{SocialAuthentication} did not implement a working prototype for social authentication.
They justified this decision with the complicated process of acquiring approval from the application providers to access social media accounts.
Instead, they evaluated a mock-up prototype without designing or implementing any protocol.
The prototype communicated with servers storing the association of social media profiles to Signal keys, simulating a perfectly secure world.
Nonetheless, the authors note that safety numbers should be posted publicly on the social media platforms, thereby disclosing the association of keys with accounts.
Through such public posts, users have no control over who can associate keys with accounts.

In contrast to both of the above works, SOAP is a privacy-preserving protocol that allows for the selective disclosure of key-to-account associations.
Whereas the proposal of \cite{SocialAuthentication} requires users to authorize Signal to access the respective social media platform, SOAP neither requires Signal to access accounts at \glspl{idp} nor vice versa.
Additionally, we demonstrated SOAP's feasibility with two fully functional prototypes, and we rigorously evaluated SOAP's security.

\paragraph{Key Authentication with OpenID Connect}
Zoom's cryptography whitepaper \cite{ZoomE2EWhitepaper} proposes ``Identity Provider Attestations,'' which authenticate Zoom's end-to-end encryption keys using OpenID Connect in conjunction with DNS.
Organizations can delegate an \gls{idp} via DNS as eligible to authenticate that organization's users.
The Zoom client can then
\begin{enumerate*}[label=(\roman*)]
  \item upload a commitment to a user's public key at the \gls{idp} on the user's behalf using OAuth and a custom API, and
  \item request an ID token that includes that commitment.
\end{enumerate*}
Both these steps require adoption by the \gls{idp}, which is not the case for SOAP.
Moreover, Zoom's design considers the case in which an account delegates authentication to a trusted \gls{idp}.
In contrast, we propose to utilize multiple \glspl{idp}, which makes it strictly more difficult to compromise a messaging account.

\subsection{Sender Correspondence}
\label{sec:rw-props}

We next focus on sender correspondence, our formalization of social authentication.
We show that this notion has application beyond secure messaging, and we relate it to existing notions of authentication.
Namely, we show that sender correspondence establishes non-injective agreement for each of the two associated pseudonyms.

\subsubsection{Sender Correspondence in the Wild}

To show that sender correspondence applies beyond social authentication as presented in this paper, consider the \gls{acme} protocol \cite{ACMERFC,LetsEncrypt}, powering the free \gls{ca} \textit{Let's Encrypt}.
\Gls{acme} automates the certificate request and issuance procedure.
Using \gls{acme}, \glspl{ca} verify certificate requests using a challenge-response mechanism in three steps.
First, a \gls{ca} receives a certificate request for a given domain name, signed by a private key.
Second, the \gls{ca} sends a challenge to the respective public keyholder and asks them to return it using DNS or HTTP.
Third, the \gls{ca} verifies that they receive the signed challenge through the DNS or HTTP channel, at which point it issues a certificate for the respective public key.

The \gls{acme} protocol can be seen as establishing sender correspondence.
Namely, the requesting public key is the pseudonym $A$ and the domain name is the pseudonym $B$.
Related literature \cite{ACMEDYStar,SeemsLegit} verifying the \gls{acme} protocol only considered key establishment properties.
Namely, they require that for any attack on \gls{acme}, the adversary must know the certificate's corresponding secret key.
However, they do not consider identity misbinding attacks, where the adversary could provide the domain name that gets associated with an honest key.
\cite{ACMERFC} analyzed \gls{acme}['s] domain validation algorithm and considered an authentication-style property.
But as the authors only analyzed this one part of \gls{acme}, they did not consider \gls{acme}'s overarching security goals.

In general, sender correspondence can be applied to any two channels that allow for information exchange associated to pseudonyms.
An \gls{idp}-managed online version-control system like GitLab is a channel where users exchange information (commits, comments, etc.) associated with a pseudonym (usernames).
Similarly, DNS is a channel where information (DNS records) is associated to pseudonyms (domain names).

\subsubsection{Relationship to Other Authentication Properties}

\paragraph{Non-Injective Agreement}
In his analysis of authentication properties \cite{AuthenticationSpec}, \Citeauthor{AuthenticationSpec} defined the notion of \textit{non-injective agreement}, which we recast slightly as follows and formalize in Tamarin's property language below:\footnote{%
  \cite{AuthenticationSpec} also requires that $A$ and $B$ agree on the intended recipient ($A$), which we drop so that we can also consider only sender-authenticated protocols.
}

\begin{authprop}[Non-injective agreement; adapted from \cite{AuthenticationSpec}]
  A protocol guarantees a responder $A$ \textit{non-injective agreement} if whenever $A$ receives a message $m$, apparently from initiator $B$, then $B$ was previously running the protocol as the initiator, and the two agents agreed on $m$.
\end{authprop}

\begin{lstlisting}[gobble=2,numbers=left]
  All R S m #tr. Receive(R, S, m) @ #tr
    ==> (Ex #ts. Send(S, m) @ #ts
          & #ts < #tr)
\end{lstlisting}

Compare this to the following formalization of sender correspondence, a simplified variant of our formalization of social authentication (Sec.~\ref{sec:analysis}).

\begin{lstlisting}[gobble=2,numbers=left,escapechar=*]
  All V R1 R2 PX PY mx my #ta #trx #try.
        ( Correspond(V, PX, PY) @ #ta
        & ReceiveChX(R1, PX, mx) @ #trx *\label{ln:ec-rx}*
        & ReceiveChY(R2, PY, my) @ #try) *\label{ln:ec-ry}*
    ==> (Ex S #tsx #tsy.
            SendChX(PX, mx) @ #tsx *\label{ln:ec-sx}*
          & Sender(S) @ #tsx
          & #tsx < #trx *\label{ln:ec-tx}*
          & SendChY(PY, my) @ #tsy *\label{ln:ec-sy}*
          & Sender(S) @ #tsy
          & #tsy < #try) *\label{ln:ec-ty}*
\end{lstlisting}

\lstinline{Receive[ChX/Y](R, S, m)} models that \lstinline{R} received message \lstinline{m} from \lstinline{S} (on channel \lstinline{X} or \lstinline{Y}), \lstinline{Send[ChX/Y](S, m)} that \lstinline{S} sent message \lstinline{m} (on channel \lstinline{X} or \lstinline{Y}).
Note that \lstinline{S} could be a pseudonym.
\lstinline{Correspond(V, PX, PY)} formalizes that the verifier \lstinline{V} identifies pseudonyms \lstinline{PX} and \lstinline{PY} with the same party, and \lstinline{Sender(S) @ #t} that the message sent at time point \lstinline{t} was sent by agent \lstinline{S}.

Intuitively, sender correspondence relates to non-injective agreement for two reasons:
\begin{enumerate*}[label=(\roman*)]
  \item Sender correspondence only works when the two associated pseudonyms can be used for authentic communication.
  Otherwise, it would make little sense to ``tie'' them together.
  \item \label{itm:formal-match} The formalizations of both properties are very similar: lines \ref{ln:ec-rx}, \ref{ln:ec-sx}, \ref{ln:ec-tx} and lines \ref{ln:ec-ry}, \ref{ln:ec-sy}, \ref{ln:ec-ty} exactly match our formalization of non-injective agreement.
\end{enumerate*}

We can express \ref{itm:formal-match} formally.
Namely, sender correspondence establishes non-injective agreement on channel \lstinline{X} and \lstinline{Y} for the pseudonyms \lstinline{PX} and \lstinline{PY}.
Specifically, we show that whenever there is a successful run of a protocol providing sender correspondence between \lstinline{PX} and \lstinline{PY}, that trace also satisfies non-injective agreement for both of these pseudonyms (formalized using the respective \lstinline{ChX} and \lstinline{ChY} events).
If such a trace were a counterexample to non-injective agreement, e.g., for channel \lstinline{X} (the other case follows symmetrically), there must be an event \lstinline{ReceiveChX(R, PX, m)} for which there is no corresponding \lstinline{SendChX} event.
In that case, since \lstinline{R1} and \lstinline{mx} in the formalization of sender correspondence are universally quantified, that trace would be a counterexample to sender correspondence as well.
This contradicts our assumption that the protocol provides sender correspondence.

This relationship between sender correspondence and non-injective agreement highlights that sender correspondence is a desirable authentication property.
When successfully running a protocol providing sender correspondence, we know that both pseudonyms can be used for authentic communication \textit{and} that they are controlled by the same sender.

\paragraph{Sender Invariance}
In \cite{SenderInv}, the authors distinguish two notions of authentication that we conflate: \textit{(non-injective) agreement} and \textit{sender invariance}.
Whereas non-injective agreement (in \cite{SenderInv}) is defined for all kinds of agents, sender invariance is defined only for pseudonyms, capturing the security guarantees behind authenticated channels that use unauthenticated public keys.
One does not know who one is connected with, but it must always be the same agent, provided their corresponding private key does not leak.
The authors show that non-injective agreement implies sender invariance.
Thus, in the formalism of \cite{SenderInv}, sender correspondence establishes both non-injective agreement and sender invariance.

\subsection{Formal Analyses of OAuth Protocols}
\label{sec:rw-analysis}

Our formal analysis of SOAP (Sec.~\ref{sec:analysis}) was influenced by the recommendations of \citetitle{OAuthSecBestPractices} standard \cite{OAuthSecBestPractices} and the formal analyses of the OAuth 2.0 and OpenID Connect protocols conducted in \cite{WIMOauth2,WIMOpenID}.
The latter works, \cite{WIMOauth2,WIMOpenID}, conducted pen-and-paper proofs in the Web Infrastructure Model \cite{WIM}, which captures more details of the browser environment than our model, e.g., HTTP status codes and their semantics.
In contrast, we utilize the Tamarin prover \cite{Tamarin}, generating machine-checked proofs and consider a strictly stronger adversary than both \cite{WIMOauth2,WIMOpenID} and the original specifications \cite{OAuthRFC,OpenIdConnect,OAuthThreatModel}.
Neither of these considered the leakage of authorization requests.

Only \cite{WIMFAPI}, the formal analysis of the \acrlong{fapi}, considers a stronger attacker model not requiring Assumption~\ref{itm:fwd-confidential} (browser redirect parameters remain confidential).
However, \cite{WIMFAPI} analyzes a different profile of OAuth 2.0 that does not match our setting as it assumes that applications can protect secrets, which enables the \gls{idp} to authenticate clients.
Dropping Assumption~\ref{itm:fwd-confidential} for SOAP would allow the adversary to capture redirects and thus effectively allow them to run the protocol themselves altogether (see Sec.~\ref{sec:discuss-threat-model}).

\cite{PriPreOIDC} proposed the \gls{poidc} protocol, which enhances OpenID Connect's privacy guarantees, and also analyzed the security of their proposal in Tamarin.
While \cite{PriPreOIDC} models the process of users granting consent more explicitly (logging in and providing consent are two steps, which we model as one), they make stronger assumptions on user behavior.
Namely, they require that users only log in and consent to OpenID Connect flows when they themselves launched the protocol.
We do not make this assumption and it is unrealistically strong.
Some \glspl{idp} neither require a login (given an existing session) nor require consent (given that consent has been granted in the past; see Sec.~\ref{apx:oidc}).
Moreover, \cite{PriPreOIDC} does not model the authorization code flow with \gls{pkce}, which our design relies upon.
Nevertheless, our design's privacy guarantees could be enhanced if designs such as \gls{poidc} were adopted.

\section{Conclusion}
\label{sec:conclusion}
Social authentication is an exciting authentication paradigm promising usable \cite{SocialAuthentication}, remote, and automated authentication in messaging applications.
In this paper, we precisely and formally defined social authentication (Sec.~\ref{sec:prop}), we presented SOAP, a secure and practical protocol implementing social authentication (Sec.~\ref{sec:design}), we formally proved that SOAP implements social authentication even in the presence of a strong adversary (Sec.~\ref{sec:security}), and we demonstrated SOAP's practicality in two prototypes (Sec.~\ref{sec:proto}).

Note that while we targeted Signal in our prototype development, and additionally WhatsApp in our problem motivation (Sec.~\ref{sec:bg}), SOAP can be applied to any application to authenticate key material and, more generally, applied to any kind of pseudonyms.
For example, the applications Telegram, Threema, and Viber \cite{TelegramContactVerification,ThreemaContactVerification,ViberContactVerification} all provide contact verification mechanisms similar to Signal and WhatsApp.
Hence, SOAP can also be applied to these applications.

SOAP is automated to a large degree and can immediately be adopted by \glspl{idp} (indeed, it may not require adoption at all) because it relies on the well-established OpenID Connect protocol.
It implements a secure and complete in-application ceremony that requires nothing more of users than their consent.
Widespread adoption in messaging applications would be a cost-effective measure to increase their robustness against impersonation attacks and eavesdropping.

\paragraph{Future Work.}
Our results suggest several next steps.
First, we argue that social authentication should be supported by modern messaging applications.
Improving our open-source Signal prototype such that it could be deployed in production is a promising first step in that direction.
Second, social authentication should be applicable far beyond secure messaging.
For example, it could be used to secure other communication such as e-mail, or video conferencing, or it could be used as a second factor.
Third, we suggest amending the OpenID Connect specification to support user-submitted claims such that users can consent unambiguously and developers can use streamlined APIs.
Finally, while \cite{SocialAuthentication} established social authentication as a usable authentication ceremony and our prototypes require little interaction, a user study would help to finalize our design, accounting for users' understanding and preferences regarding social authentication.

\if\anonymize0
\section*{Acknowledgements}
This research was funded by the Werner Siemens-Stiftung (WSS) as part of the Centre for Cyber Trust (CECYT).
We thank the WSS for their contribution.
We also thank our colleagues Mihael Liskij and Jorge Luis Toro Pozo for their valuable feedback in the early stages of this project.
\fi

\printbibliography

@inproceedings{ACMEDYStar,
  title = {An {{In-Depth Symbolic Security Analysis}} of the {{ACME Standard}}},
  booktitle = {Proceedings of the 2021 {{ACM SIGSAC Conference}} on {{Computer}} and {{Communications Security}}},
  author = {Bhargavan, Karthikeyan and Bichhawat, Abhishek and Do, Quoc Huy and Hosseyni, Pedram and Küsters, Ralf and Schmitz, Guido and Würtele, Tim},
  date = {2021-11-12},
  series = {{{CCS}} '21},
  pages = {2601--2617},
  publisher = {{ACM}},
  location = {{New York, NY, USA}},
  doi = {10.1145/3460120.3484588},
  isbn = {978-1-4503-8454-4}
}

@report{ACMERFC,
  type = {Request for Comments},
  title = {Automatic {{Certificate Management Environment}} ({{ACME}})},
  author = {Barnes, Richard and Hoffman-Andrews, Jacob and McCarney, Daniel and Kasten, James},
  date = {2019-03},
  number = {RFC 8555},
  institution = {{Internet Engineering Task Force}},
  doi = {10.17487/RFC8555},
  pagetotal = {95}
}

@inproceedings{ActionNeeded,
  title = {Action {{Needed}}! {{Helping Users Find}} and {{Complete}} the {{Authentication Ceremony}} in {{Signal}}},
  author = {Vaziripour, Elham and Wu, Justin and O'Neill, Mark and Metro, Daniel and Cockrell, Josh and Moffett, Timothy and Whitehead, Jordan and Bonner, Nick and Seamons, Kent and Zappala, Daniel},
  date = {2018},
  pages = {47--62},
  url = {https://www.usenix.org/conference/soups2018/presentation/vaziripour},
  urldate = {2022-07-18},
  eventtitle = {Fourteenth {{Symposium}} on {{Usable Privacy}} and {{Security}} ({{SOUPS}} 2018)},
  isbn = {978-1-939133-10-6},
  langid = {english}
}

@online{AndroidIntentsChrome,
  title = {Android {{Intents}} with {{Chrome}}},
  url = {https://developer.chrome.com/docs/multidevice/android/intents/},
  urldate = {2022-10-03},
  langid = {english},
  organization = {{Chrome Developers}}
}

@online{AndroidLinks,
  title = {Handling {{Android App Links}}},
  url = {https://developer.android.com/training/app-links},
  urldate = {2022-07-15},
  langid = {english},
  organization = {{Android Developers}}
}

@inproceedings{AnonymousKeyTransparency,
  title = {Automatic {{Detection}} of {{Fake Key Attacks}} in {{Secure Messaging}}},
  booktitle = {Proceedings of the 2022 {{ACM SIGSAC Conference}} on {{Computer}} and {{Communications Security}}},
  author = {Yadav, Tarun Kumar and Gosain, Devashish and Herzberg, Amir and Zappala, Daniel and Seamons, Kent},
  date = {2022-11-07},
  series = {{{CCS}} '22},
  pages = {3019--3032},
  publisher = {{Association for Computing Machinery}},
  location = {{New York, NY, USA}},
  doi = {10.1145/3548606.3560588},
  isbn = {978-1-4503-9450-5}
}

@inproceedings{AuthenticationSpec,
  title = {A Hierarchy of Authentication Specifications},
  booktitle = {Proceedings 10th {{Computer Security Foundations Workshop}}},
  author = {Lowe, G.},
  date = {1997-06},
  pages = {31--43},
  doi = {10.1109/CSFW.1997.596782},
  eventtitle = {Proceedings 10th {{Computer Security Foundations Workshop}}}
}

@inproceedings{Coniks,
  title = {{{CONIKS}}: {{Bringing Key Transparency}} to {{End Users}}},
  shorttitle = {{{CONIKS}}},
  booktitle = {24th {{USENIX Security Symposium}}},
  author = {Melara, Marcela S. and Blankstein, Aaron and Bonneau, Joseph and Felten, Edward W. and Freedman, Michael J.},
  date = {2015},
  pages = {383--398},
  publisher = {{USENIX Association}},
  location = {{Washington, D.C., USA}},
  url = {https://www.usenix.org/conference/usenixsecurity15/technical-sessions/presentation/melara},
  urldate = {2021-09-24},
  eventtitle = {{{USENIX Security}} 15},
  isbn = {978-1-939133-11-3},
  langid = {english}
}

@inproceedings{E2EAuthUserStudy1,
  title = {When {{SIGNAL}} Hits the {{Fan}}: {{On}} the {{Usability}} and {{Security}} of {{State-of-the-Art Secure Mobile Messaging}} – {{NDSS Symposium}}},
  booktitle = {23rd {{Annual Network}} and {{Distributed System Security Symposium}}},
  author = {Schröder, Svenja and Huber, Markus and Wind, David and Rottermanner, Christoph},
  date = {2016-08-12},
  series = {{{NDSS}} 2016},
  publisher = {{The Internet Society}},
  location = {{San Diego, California, USA}},
  url = {https://www.ndss-symposium.org/ndss2016/eurousec-2016-workshop/when-signal-hits-fan-usability-and-security-state-art-secure-mobile-messaging/},
  urldate = {2022-04-19},
  langid = {american}
}

@inproceedings{E2EAuthUserStudy2,
  title = {Is That You, {{Alice}}? {{A Usability Study}} of the {{Authentication Ceremony}} of {{Secure Messaging Applications}}},
  shorttitle = {Is That You, {{Alice}}?},
  author = {Vaziripour, Elham and Wu, Justin and O'Neill, Mark and Whitehead, Jordan and Heidbrink, Scott and Seamons, Kent and Zappala, Daniel},
  date = {2017},
  pages = {29--47},
  url = {https://www.usenix.org/conference/soups2017/technical-sessions/presentation/vaziripour},
  urldate = {2022-04-19},
  eventtitle = {Thirteenth {{Symposium}} on {{Usable Privacy}} and {{Security}} ({{SOUPS}} 2017)},
  isbn = {978-1-931971-39-3},
  langid = {english}
}

@inproceedings{E2EAuthUserStudy3,
  title = {Can {{Johnny}} Finally Encrypt? Evaluating {{E2E-encryption}} in Popular {{IM}} Applications},
  shorttitle = {Can {{Johnny}} Finally Encrypt?},
  booktitle = {Proceedings of the 6th {{Workshop}} on {{Socio-Technical Aspects}} in {{Security}} and {{Trust}}},
  author = {Herzberg, Amir and Leibowitz, Hemi},
  date = {2016-12-05},
  series = {{{STAST}} '16},
  pages = {17--28},
  publisher = {{ACM}},
  location = {{New York, NY, USA}},
  doi = {10.1145/3046055.3046059},
  isbn = {978-1-4503-4826-3}
}

@online{GoogleSignInSDKAndroid,
  title = {Start {{Integrating Google Sign-In}} into {{Your Android App}} | {{Google Sign-In}} for {{Android}} | {{Google Developers}}},
  url = {https://developers.google.com/identity/sign-in/android/start-integrating},
  urldate = {2022-10-04},
  langid = {english}
}

@inproceedings{InteractiveSignalVerification,
  title = {A {{Formal Security Analysis}} of the {{Signal Messaging Protocol}}},
  booktitle = {2017 {{IEEE European Symposium}} on {{Security}} and {{Privacy}}},
  author = {Cohn-Gordon, Katriel and Cremers, Cas and Dowling, Benjamin and Garratt, Luke and Stebila, Douglas},
  date = {2017-04},
  pages = {451--466},
  doi = {10.1109/EuroSP.2017.27},
  eventtitle = {{{EuroS}}\&{{P}}}
}

@report{JSON,
  type = {Request for Comments},
  title = {The {{JavaScript Object Notation}} ({{JSON}}) {{Data Interchange Format}}},
  author = {Bray, Tim},
  date = {2014-03},
  number = {RFC 7159},
  institution = {{Internet Engineering Task Force}},
  doi = {10.17487/RFC7159},
  pagetotal = {16}
}

@report{JWS,
  type = {Request for Comments},
  title = {{{JSON Web Signature}} ({{JWS}})},
  author = {Jones, Michael and Bradley, John and Sakimura, Nat},
  date = {2015-05},
  number = {RFC 7515},
  institution = {{Internet Engineering Task Force}},
  doi = {10.17487/RFC7515},
  pagetotal = {59}
}

@online{Keybase,
  title = {Keybase},
  url = {https://keybase.io/},
  urldate = {2022-07-18}
}

@online{KeybaseDocs,
  title = {Keybase {{Book}}: {{Learn}} about Your {{Keybase}} Account},
  url = {https://book.keybase.io/account#proofs},
  urldate = {2022-07-18}
}

@inproceedings{LetsEncrypt,
  title = {Let's {{Encrypt}}: {{An Automated Certificate Authority}} to {{Encrypt}} the {{Entire Web}}},
  shorttitle = {Let's {{Encrypt}}},
  booktitle = {Proceedings of the 2019 {{ACM SIGSAC Conference}} on {{Computer}} and {{Communications Security}}},
  author = {Aas, Josh and Barnes, Richard and Case, Benton and Durumeric, Zakir and Eckersley, Peter and Flores-López, Alan and Halderman, J. Alex and Hoffman-Andrews, Jacob and Kasten, James and Rescorla, Eric and Schoen, Seth and Warren, Brad},
  date = {2019-11-06},
  series = {{{CCS}} '19},
  pages = {2473--2487},
  publisher = {{ACM}},
  location = {{New York, NY, USA}},
  doi = {10.1145/3319535.3363192},
  isbn = {978-1-4503-6747-9}
}

@report{OAuthRFC,
  type = {Request for Comments},
  title = {The {{OAuth}} 2.0 {{Authorization Framework}}},
  author = {Hardt, Dick},
  date = {2012-10},
  number = {RFC 6749},
  institution = {{Internet Engineering Task Force}},
  doi = {10.17487/RFC6749},
  pagetotal = {76}
}

@report{OAuthSecBestPractices,
  type = {Internet Draft},
  title = {{{OAuth}} 2.0 {{Security Best Current Practice}}},
  author = {Lodderstedt, Torsten and Bradley, John and Labunets, Aney and Fett, Daniel},
  date = {2021-12-16},
  number = {draft-ietf-oauth-security-topics-19},
  institution = {{Internet Engineering Task Force}},
  url = {https://www.ietf.org/archive/id/draft-ietf-oauth-security-topics-24.html},
  urldate = {2023-12-22},
  pagetotal = {52}
}

@report{OAuthThreatModel,
  type = {Request for Comments},
  title = {{{OAuth}} 2.0 {{Threat Model}} and {{Security Considerations}}},
  author = {Lodderstedt, Torsten and McGloin, Mark and Hunt, Phil},
  date = {2013-01},
  number = {RFC 6819},
  institution = {{Internet Engineering Task Force}},
  doi = {10.17487/RFC6819},
  pagetotal = {71}
}

@report{OIDCDiscovery,
  title = {{{OpenID Connect Discovery}} 1.0 Incorporating Errata Set 1},
  author = {Sakimura, N. and Bradley, J. and Jones, M. and Jay, E.},
  date = {2014-11-08},
  url = {https://openid.net/specs/openid-connect-discovery-1_0.html},
  urldate = {2022-07-15}
}

@report{OpenIdConnect,
  title = {{{OpenID Connect Core}} 1.0 Incorporating Errata Set 1},
  author = {Sakimura, N. and Bradley, J. and Jones, M. and family=Medeiros, given=B., prefix=de, useprefix=true and Mortimore, C.},
  date = {2014-11-08},
  url = {https://openid.net/specs/openid-connect-core-1_0.html},
  urldate = {2021-09-24}
}

@inproceedings{Parakeet,
  title = {Parakeet: {{Practical Key Transparency}} for {{End-to-End Encrypted Messaging}}},
  shorttitle = {Parakeet},
  booktitle = {30th {{Annual Network}} and {{Distributed System Security Symposium}}},
  author = {Malvai, Harjasleen and Kokoris-Kogias, Lefteris and Sonnino, Alberto and Ghosh, Esha and Oztürk, Ercan and Lewi, Kevin and Lawlor, Sean F.},
  date = {0027-02/2023-03-03},
  publisher = {{The Internet Society}},
  location = {{San Diego, California, USA}},
  url = {https://www.ndss-symposium.org/ndss-paper/parakeet-practical-key-transparency-for-end-to-end-encrypted-messaging/},
  urldate = {2023-04-24},
  eventtitle = {{{NDSS}} 2023}
}

@report{PKCE,
  type = {Request for Comments},
  title = {Proof {{Key}} for {{Code Exchange}} by {{OAuth Public Clients}}},
  author = {Sakimura, Nat and Bradley, John and Agarwal, Naveen},
  date = {2015-09},
  number = {RFC 7636},
  institution = {{Internet Engineering Task Force}},
  doi = {10.17487/RFC7636}
}

@inproceedings{PriPreOIDC,
  title = {Privacy-{{Preserving OpenID Connect}}},
  booktitle = {Proceedings of the 15th {{ACM Asia Conference}} on {{Computer}} and {{Communications Security}}},
  author = {Hammann, Sven and Sasse, Ralf and Basin, David},
  date = {2020-10-05},
  series = {{{ASIA CCS}} '20},
  pages = {277--289},
  publisher = {{ACM}},
  location = {{New York, NY, USA}},
  doi = {10.1145/3320269.3384724},
  isbn = {978-1-4503-6750-9}
}

@software{React,
  title = {React},
  date = {2022-07-19T15:35:43Z},
  origdate = {2013-05-24T16:15:54Z},
  url = {https://github.com/facebook/react},
  urldate = {2022-07-19},
  organization = {{Meta}}
}

@inproceedings{RemoteFingerprintComp,
  title = {On the {{Pitfalls}} of {{End-to-End Encrypted Communications}}: {{A Study}} of {{Remote Key-Fingerprint Verification}}},
  shorttitle = {On the {{Pitfalls}} of {{End-to-End Encrypted Communications}}},
  booktitle = {Proceedings of the 33rd {{Annual Computer Security Applications Conference}}},
  author = {Shirvanian, Maliheh and Saxena, Nitesh and George, Jesvin James},
  date = {2017-12-04},
  series = {{{ACSAC}} '17},
  pages = {499--511},
  publisher = {{ACM}},
  location = {{New York, NY, USA}},
  doi = {10.1145/3134600.3134610},
  isbn = {978-1-4503-5345-8}
}

@inproceedings{SEEMless,
  title = {{{SEEMless}}: {{Secure End-to-End Encrypted Messaging}} with Less {{Trust}}},
  shorttitle = {{{SEEMless}}},
  booktitle = {Proceedings of the 2019 {{ACM SIGSAC Conference}} on {{Computer}} and {{Communications Security}}},
  author = {Chase, Melissa and Deshpande, Apoorvaa and Ghosh, Esha and Malvai, Harjasleen},
  editor = {Cavallaro, Lorenzo and Kinder, Johannes and Wang, XiaoFeng and Katz, Jonathan},
  date = {2019-11-11/2019-11-15},
  pages = {1639--1656},
  publisher = {{ACM}},
  location = {{London, UK}},
  doi = {10.1145/3319535.3363202},
  eventtitle = {{{CCS}} 2019}
}

@inproceedings{SeemsLegit,
  title = {Seems {{Legit}}: {{Automated Analysis}} of {{Subtle Attacks}} on {{Protocols}} That {{Use Signatures}}},
  shorttitle = {Seems {{Legit}}},
  booktitle = {Proceedings of the 2019 {{ACM SIGSAC Conference}} on {{Computer}} and {{Communications Security}}},
  author = {Jackson, Dennis and Cremers, Cas and Cohn-Gordon, Katriel and Sasse, Ralf},
  date = {2019-11-06},
  series = {{{CCS}} '19},
  pages = {2165--2180},
  publisher = {{ACM}},
  location = {{New York, NY, USA}},
  doi = {10.1145/3319535.3339813},
  isbn = {978-1-4503-6747-9}
}

@inproceedings{SenderInv,
  title = {Formalizing and Analyzing Sender Invariance},
  booktitle = {Proceedings of the 4th International Conference on {{Formal}} Aspects in Security and Trust},
  author = {Drielsma, Paul Hankes and Mödersheim, Sebastian and Viganò, Luca and Basin, David},
  date = {2006-08-26},
  series = {{{FAST}}'06},
  pages = {80--95},
  publisher = {{Springer-Verlag}},
  location = {{Berlin, Heidelberg}},
  isbn = {978-3-540-75226-4}
}

@online{SignalOTPCompromise,
  title = {Twilio {{Incident}}: {{What Signal Users Need}} to {{Know}}},
  shorttitle = {Twilio {{Incident}}},
  url = {https://support.signal.org/hc/en-us/articles/4850133017242-Twilio-Incident-What-Signal-Users-Need-to-Know-},
  urldate = {2022-08-30},
  langid = {american},
  organization = {{Signal Support}}
}

@online{SignalOTPCompromiseVictim,
  title = {How a {{Third-Party SMS Service Was Used}} to {{Take Over Signal Accounts}}},
  author = {Franceschi-Bicchierai, Lorenzo},
  date = {2022-08-17T19:27:21},
  url = {https://www.vice.com/en/article/qjkvxv/how-a-third-party-sms-service-was-used-to-take-over-signal-accounts},
  urldate = {2022-08-30},
  langid = {english},
  organization = {{Vice}}
}

@online{SignalRegistrationLock,
  title = {Signal {{PIN}}},
  url = {https://support.signal.org/hc/en-us/articles/360007059792-Signal-PIN},
  urldate = {2023-04-06},
  langid = {american},
  organization = {{Signal Support}}
}

@online{SignalSafetyNumbers,
  title = {Safety Number Updates},
  author = {{moxie0}},
  url = {https://signal.org/blog/safety-number-updates/},
  urldate = {2022-01-17},
  langid = {english},
  organization = {{Signal Messenger}}
}

@online{SignalWhatsAppE2E,
  title = {{{WhatsApp}}'s {{Signal Protocol}} Integration Is Now Complete},
  url = {https://signal.org/blog/whatsapp-complete/},
  urldate = {2022-08-04},
  langid = {english},
  organization = {{Signal Messenger}}
}

@article{SMSAuthInsecure,
  title = {Security {{Analysis}} of {{SMS}} as a {{Second Factor}} of {{Authentication}}: {{The}} Challenges of Multifactor Authentication Based on {{SMS}}, Including Cellular Security Deficiencies, {{SS7}} Exploits, and {{SIM}} Swapping},
  shorttitle = {Security {{Analysis}} of {{SMS}} as a {{Second Factor}} of {{Authentication}}},
  author = {Jover, Roger Piqueras},
  date = {2020-08-31},
  journaltitle = {Queue},
  shortjournal = {Queue},
  volume = {18},
  number = {4},
  pages = {Pages 20:37--Pages 20:60},
  issn = {1542-7730},
  doi = {10.1145/3424302.3425909}
}

@inproceedings{SocialAuthentication,
  title = {I {{Don}}'t {{Even Have}} to {{Bother Them}}!: {{Using Social Media}} to {{Automate}} the {{Authentication Ceremony}} in {{Secure Messaging}}},
  booktitle = {Proceedings of the 2019 {{CHI Conference}} on {{Human Factors}} in {{Computing Systems}}},
  author = {Vaziripour, Elham and Howard, Devon and Tyler, Jake and O'Neill, Mark and Wu, Justin and Seamons, Kent and Zappala, Daniel},
  date = {2019-05},
  location = {{New York, NY, USA}},
  url = {https://doi.org/10.1145/3290605.3300323},
  urldate = {2022-06-16},
  eventtitle = {{{CHI}} '19},
  langid = {english}
}

@inproceedings{Tamarin,
  title = {The {{TAMARIN Prover}} for the {{Symbolic Analysis}} of {{Security Protocols}}},
  booktitle = {Computer {{Aided Verification}}},
  author = {Meier, Simon and Schmidt, Benedikt and Cremers, Cas and Basin, David},
  editor = {Sharygina, Natasha and Veith, Helmut},
  date = {2013},
  series = {Lecture {{Notes}} in {{Computer Science}}},
  pages = {696--701},
  publisher = {{Springer}},
  location = {{Berlin, Heidelberg}},
  doi = {10.1007/978-3-642-39799-8_48},
  isbn = {978-3-642-39799-8},
  langid = {english}
}

@inproceedings{Tamarin5GAKA,
  title = {Component-Based Formal Analysis of {{5G-AKA}}: Channel Assumptions and Session Confusion},
  shorttitle = {Component-Based Formal Analysis of {{5G-AKA}}},
  booktitle = {26th {{Annual Network}} and {{Distributed System Security Symposium}}},
  author = {Cremers, C. and Dehnel-Wild, M.},
  date = {2019},
  series = {{{NDSS}} 2019},
  publisher = {{The Internet Society}},
  url = {https://www.ndss-symposium.org/ndss-paper/component-based-formal-analysis-of-5g-aka-channel-assumptions-and-session-confusion/},
  urldate = {2022-10-10},
  langid = {english}
}

@inproceedings{TamarinEMV,
  title = {The {{EMV Standard}}: {{Break}}, {{Fix}}, {{Verify}}},
  shorttitle = {The {{EMV Standard}}},
  booktitle = {2021 {{IEEE Symposium}} on {{Security}} and {{Privacy}} ({{S}}\&{{P}})},
  author = {Basin, David and Sasse, Ralf and Toro-Pozo, Jorge},
  date = {2021-05},
  pages = {1766--1781},
  issn = {2375-1207},
  doi = {10.1109/SP40001.2021.00037}
}

@inproceedings{TamarinTLS,
  title = {A {{Comprehensive Symbolic Analysis}} of {{TLS}} 1.3},
  booktitle = {Proceedings of the 2017 {{ACM SIGSAC Conference}} on {{Computer}} and {{Communications Security}}},
  author = {Cremers, Cas and Horvat, Marko and Hoyland, Jonathan and Scott, Sam and family=Merwe, given=Thyla, prefix=van der, useprefix=true},
  date = {2017-10-30},
  series = {{{CCS}} '17},
  pages = {1773--1788},
  publisher = {{ACM}},
  location = {{New York, NY, USA}},
  doi = {10.1145/3133956.3134063},
  isbn = {978-1-4503-4946-8}
}

@online{TelegramContactVerification,
  title = {Telegram {{FAQ}}},
  url = {https://telegram.org/faq?setln=en#q-what-is-this-39encryption-key-39-thing},
  urldate = {2023-12-19},
  organization = {{Telegram}}
}

@online{ThreemaContactVerification,
  title = {What Do the Three Colored Dots next to a Contact Mean? – {{Threema}}},
  shorttitle = {What Do the Three Colored Dots next to a Contact Mean?},
  url = {https://threema.ch/en/faq/levels_expl},
  urldate = {2023-12-19},
  langid = {english}
}

@report{TLS,
  type = {Request for Comments},
  title = {The {{Transport Layer Security}} ({{TLS}}) {{Protocol Version}} 1.3},
  author = {Rescorla, Eric},
  date = {2018-08},
  number = {RFC 8446},
  institution = {{Internet Engineering Task Force}},
  doi = {10.17487/RFC8446},
  pagetotal = {160}
}

@inproceedings{UserCentricDesignSignal,
  title = {Exploring {{User-Centered Security Design}} for {{Usable Authentication Ceremonies}}},
  booktitle = {Proceedings of the 2021 {{CHI Conference}} on {{Human Factors}} in {{Computing Systems}}},
  author = {Fassl, Matthias and Gröber, Lea Theresa and Krombholz, Katharina},
  date = {2021-05-06},
  series = {{{CHI}} '21},
  pages = {1--15},
  publisher = {{ACM}},
  location = {{New York, NY, USA}},
  doi = {10.1145/3411764.3445164},
  isbn = {978-1-4503-8096-6}
}

@online{ViberContactVerification,
  title = {Verify {{End-To-End Encryption}}: {{Trusted Contacts List}}},
  shorttitle = {Verify {{End-To-End Encryption}}},
  url = {https://help.viber.com/hc/en-us/articles/9061180581661-Verify-End-To-End-Encryption-Trusted-Contacts-List},
  urldate = {2023-12-19},
  langid = {american},
  organization = {{Viber}}
}

@online{WhatsAppKeyTransparency,
  title = {Deploying Key Transparency at {{WhatsApp}}},
  author = {Lewi, Kevin, Sean Lawlor},
  date = {2023-04-13T12:59:32+00:00},
  url = {https://engineering.fb.com/2023/04/13/security/whatsapp-key-transparency/},
  urldate = {2023-04-24},
  langid = {american},
  organization = {{Engineering at Meta}}
}

@software{WhatsAppKeyTransparencyCode,
  title = {Facebook/Akd},
  date = {2023-04-21T15:32:54Z},
  origdate = {2021-06-29T20:33:21Z},
  url = {https://github.com/facebook/akd},
  urldate = {2023-04-24},
  organization = {{Meta}}
}

@online{WhatsAppRegistrationLock,
  title = {About Registration and Two-Step Verification | {{WhatsApp Help Center}}},
  url = {https://faq.whatsapp.com/506595211487528/?helpref=hc_fnav},
  urldate = {2023-04-06}
}

@inproceedings{WIM,
  title = {An {{Expressive Model}} for the {{Web Infrastructure}}: {{Definition}} and {{Application}} to the {{Browser ID SSO System}}},
  shorttitle = {An {{Expressive Model}} for the {{Web Infrastructure}}},
  booktitle = {2014 {{IEEE Symposium}} on {{Security}} and {{Privacy}}},
  author = {Fett, Daniel and Küsters, Ralf and Schmitz, Guido},
  date = {2014-05},
  pages = {673--688},
  issn = {2375-1207},
  doi = {10.1109/SP.2014.49}
}

@inproceedings{WIMFAPI,
  title = {An {{Extensive Formal Security Analysis}} of the {{OpenID Financial-Grade API}}},
  booktitle = {2019 {{IEEE Symposium}} on {{Security}} and {{Privacy}} ({{S}}\&{{P}})},
  author = {Fett, Daniel and Hosseyni, Pedram and Küsters, Ralf},
  date = {2019-05},
  pages = {453--471},
  issn = {2375-1207},
  doi = {10.1109/SP.2019.00067}
}

@inproceedings{WIMOauth2,
  title = {A {{Comprehensive Formal Security Analysis}} of {{OAuth}} 2.0},
  booktitle = {Proceedings of the 2016 {{ACM SIGSAC Conference}} on {{Computer}} and {{Communications Security}}},
  author = {Fett, Daniel and Küsters, Ralf and Schmitz, Guido},
  date = {2016-10-24},
  series = {{{CCS}} '16},
  pages = {1204--1215},
  publisher = {{ACM}},
  location = {{New York, NY, USA}},
  doi = {10.1145/2976749.2978385},
  isbn = {978-1-4503-4139-4}
}

@inproceedings{WIMOpenID,
  title = {The {{Web SSO Standard OpenID Connect}}: {{In-depth Formal Security Analysis}} and {{Security Guidelines}}},
  shorttitle = {The {{Web SSO Standard OpenID Connect}}},
  booktitle = {2017 {{IEEE}} 30th {{Computer Security Foundations Symposium}} ({{CSF}})},
  author = {Fett, Daniel and Küsters, Ralf and Schmitz, Guido},
  date = {2017-08},
  pages = {189--202},
  issn = {2374-8303},
  doi = {10.1109/CSF.2017.20},
  eventtitle = {2017 {{IEEE}} 30th {{Computer Security Foundations Symposium}} ({{CSF}})}
}

@online{WindowsLinks,
  title = {Enable Apps for Websites Using App {{URI}} Handlers},
  url = {https://docs.microsoft.com/en-us/windows/uwp/launch-resume/web-to-app-linking},
  urldate = {2022-07-15},
  langid = {american}
}

@misc{ZoomE2EWhitepaper,
  title = {Zoom {{Cryptography Whitepaper}}},
  author = {Blum, Josh and Booth, Simon and Chen, Brian and Gal, Oded and Krohn, Maxwell and Len, Julia and Lyons, Karan and Marcedone, Antonio and Maxim, Mike and Mou, Marry Ember and Namavari, Armin and O'Connor, Jack and Rien, Surya and Steele, Miles and Green, Matthew and Kissner, Lea and Stamos, Alex},
  date = {2023-11-21},
  url = {https://github.com/zoom/zoom-e2e-whitepaper/blob/v4.3/zoom_e2e.pdf},
  urldate = {2023-12-05},
  organization = {{Zoom Video Communications, Inc.}}
}

\appendix

\section{Detailed Protocol Description}
\label{apx:protocol}
To run SOAP, applications must periodically load the \acrshortpl{idp}'s token and authorization endpoints, as well as signing keys as specified in \cite{OIDCDiscovery}.
The simplest option to do so is whenever a application engages in SOAP.

The prover's application starts SOAP by generating three random values: a code verifier $cv$ as per the \gls{pkce} specification \cite{PKCE}, a salt $s$, and a nonce $n$.
We suggest generating each of these random values with at least 256-bits of entropy, following \cite{PKCE}'s security requirements.
The application then uses a secure password-hashing algorithm $h$ to calculate a salted hash $h(k, s)$ of the safety number $k$.
Next, the application uses these values to issue an OpenID Connect authentication request to the selected \gls{idp} with the following parameters:
\begin{description}
  \item[scope:] ``openid email''; depending on the \gls{idp}, other scopes than ``email'' may be desirable.
  \item[response\_type:] ``code''
  \item[nonce:] $n \mid\mid h(k, s)$; the application must ensure that it does not include the salt.
  $\mid\mid$ denotes concatenation.
  The application must ensure the parsing is unambiguous, e.g., by adding a delimiter character.
  \item[state:] $n$
  \item[code\_challenge:] $\text{S256}(cv)$; S256 marks the SHA-256 hashing algorithm.
  \item[code\_challenge\_method:] ``S256''
\end{description}

Naturally, the application also includes its \gls{idp}-issued application ID, and an appropriate redirect URL.
Redirect URLs must use the HTTPS scheme and must be distinct per \gls{idp}.
The application stores the salted hash $h(k ,s)$, the salt $s$, the nonce $n$, the redirect URL used, and the code verifier $cv$ as the most recently issued request.
Then, the application launches the system's browser with the request URL, which in turn takes the user to the consent and login page.

When the user consents, the \gls{idp} redirects the browser back to the application.
The application verifies that it received the authorization code through the expected redirect URL and with the expected state by comparing these values to those stored as most recently issued request.
If both checks pass, the application uses the authorization code and stored code verifier to request the ID token from the \gls{idp} as specified in \cite{OpenIdConnect}, i.e., using a POST request.
When receiving the token in the response, the application verifies the token as follows:
\begin{enumerate}
  \item Verify that the issuer matches the redirect URL stored.
  \item Verify that the token's audience matches the application ID.
  \item Verify that the token's nonce includes the hash stored.
  \item \label{itm:expire-check} Verify that the token is not expired.
  \item \label{itm:sig-check} Verify the token's signature using a key loaded from the \acrshort{idp}'s discovery document.
\end{enumerate}

If all checks pass, the application clears its storage for the most recently issued request, stores the nonce as issued by itself, and forwards the token and salt to the verifier.
The verifier's application applies checks \ref{itm:expire-check} and \ref{itm:sig-check} and makes the following additional checks:
\begin{enumerate}
  \item Verify that the nonce encoded in the token has not been generated by itself, comparing it to the nonces stored locally.
  \item Verify the safety number's hash encoded in the token by recomputing it, using the salt provided by the prover and the safety number of the channel through which it received the token, and comparing it for equality.
\end{enumerate}

\end{document}